\documentclass[11pt,draftcls,onecolumn]{IEEEtran}

\usepackage{amsfonts}
\usepackage{amssymb}
\usepackage{mathrsfs}
\usepackage{amsmath}
\usepackage{cite}
\usepackage{mathrsfs}
\usepackage[ruled,vlined]{algorithm2e}
\usepackage{tikz}
\usetikzlibrary{arrows,automata}
\usepackage[latin1]{inputenc}
\usepackage{verbatim}
\makeatletter

\newcommand{\Rmnum}[1]{\expandafter\@slowromancap\romannumeral #1@}
\makeatother

\newtheorem{thm}{Theorem}
\newtheorem{lemma}[thm]{Lemma}

\newtheorem{cor}[thm]{Corollary}
\newtheorem{defn}{Definition}
\newtheorem{rem}[thm]{Remark}

\newtheorem{rem-eg}[thm]{Remark and Example}

\newcommand{\ti}{\tilde}

\newcommand{\w}{{\omega}}
\newcommand{\p}{{\rho}}
\newcommand{\bX}{{\bf X}}
\newcommand{\bZ}{{\bf Z}}
\newcommand{\bU}{{\bf U}}
\newcommand{\bzero}{{\bf 0}}
\newcommand{\de}{{\delta}}
\newcommand{\dt}{{\delta}}
\newcommand{\bt}{{\beta_t}}
\newcommand{\f}{\tilde{f}}
\newcommand{\g}{\tilde{g}}
\newcommand{\F}{\tilde{F}}
\newcommand{\mL}{\mathcal{L}}
\newcommand{\mF}{\mathcal{F}}
\newcommand{\mP}{\mathcal{P}}
\newcommand{\mO}{\mathcal{O}}
\newcommand{\mE}{\mathcal{E}}
\newcommand{\mT}{\mathcal{T}}
\newcommand{\Rank}{{\mathrm{Rank}}}

\newcommand{\rt}{{\rm row}_t}
\newcommand{\row}{{\rm row}}
\newcommand{\ul}{\underline}
\SetKw{KwInitialization}{Initialization:}

\begin{document}

\title{Linear Network Error Correction Multicast/Broadcast/Dispersion/Generic Codes
\thanks{This research is supported by the National Key Basic Research Program of China (973 Program Grant No. 2013CB834204), the National Natural Science Foundation of China (Nos. 61301137, 61171082, 10990011), and Fundamental Research Funds for the Central Universities of China (No. 65121007). }}

\author{Xuan~Guang, and~Fang-Wei~Fu
\thanks{X. Guang is with the School of Mathematical Sciences and LPMC, Nankai University, Tianjin 300071, P.R. China. Email:
xguang@nankai.edu.cn.}
\thanks{F.-W. Fu is with the Chern Institute of
Mathematics and LPMC, Nankai University, Tianjin 300071, P.R. China. Email:
fwfu@nankai.edu.cn.}
}

\maketitle

\begin{abstract}
In the practical network communications, many internal nodes in the network are required to not only transmit messages but decode source messages.  For different applications, four important classes of linear network codes in network coding theory, i.e., linear multicast, linear broadcast, linear dispersion, and generic network codes, have been studied extensively. More generally, when channels of communication networks are noisy,  information transmission and error correction have to be under consideration simultaneously, and thus these four classes of linear network codes are generalized to linear network error correction (LNEC) coding, and we say them LNEC multicast, broadcast, dispersion, and generic codes, respectively. Furthermore, in order to characterize their efficiency of information transmission and error correction, we propose the (weakly, strongly) extended Singleton bounds for them, and define the corresponding optimal codes, i.e., LNEC multicast/broadcast/dispersion/generic MDS codes, which satisfy the corresponding Singleton bounds with equality. The existences of such MDS codes are discussed in detail by algebraic methods and the constructive algorithms are also proposed.
\end{abstract}


\IEEEpeerreviewmaketitle

\section{Introduction}

\IEEEPARstart{N}{etwork} coding, first studied by Yeung and Zhang \cite{Zhang-Yeung-1999} and then fully developed by Ahlswede \textit{et al.} \cite{Ahlswede-Cai-Li-Yeung-2000}, reveals that if coding is applied at the nodes in a network, rather than routing alone, the source node can multicast the information to all sink nodes at the theoretically maximum rate as the alphabet size approaches infinity, where the theoretically maximum rate is the smallest minimum cut capacity between the source node and any sink node. Li \emph{et al.} \cite{Li-Yeung-Cai-2003} further showed that linear network coding with finite alphabet size is sufficient for this multicast. Koetter and M$\acute{\textup{e}}$dard \cite{Koetter-Medard-algebraic} developed an algebraic characterization for linear network coding. Although network coding increases the network capacity, Jaggi \textit{et al.} \cite{co-construction} still proposed a deterministic polynomial-time algorithm for constructing a linear network code. For a detail and comprehensive discussion of network coding, refer to \cite{Zhang-book, Yeung-book, Fragouli-book, Ho-book}.

Network coding has been studied extensively for several years under the assumption that the channels of networks are error-free. Unfortunately, all kinds of errors may occur during network communications. In order to deal with such problems, network error correction coding was studied recently. Cai and Yeung proposed the original idea of network error correction coding in their conference paper \cite{Yeung-Cai-coorrect} and developed it in their recent journal papers \cite{Yeung-Cai-correct-1}\cite{Yeung-Cai-correct-2}. They introduced the concept of network error correction codes as a generalization of classical error-correcting codes, and generalized some important bounds in classical error-correcting codes to network error correction codes, such as the Singleton bound, the Hamming bound, and the Gilbert-Varshamov bound. Although the Singleton bound was given by Yeung and Cai \cite{Yeung-Cai-correct-1}, Zhang\cite{zhang-correction} and Yang \textit{et al.} \cite{Yang-refined-Singleton}\cite{Yang-thesis} presented the refined Singleton bound independently by the different methods.
In \cite{Koetter-correction}, Koetter and Kschischang formulated a different framework for network error correction coding when a noncoherent network model was under consideration, where neither the source node nor sink nodes were assumed to have knowledge of the channel transfer characteristic. In other words, the error control problem in random linear network coding was considered. Motivated by the property that linear network coding is vector-space preserving, the source information in their approach is represented by a subspace of a fixed vector space and a basis of this subspace is injected into the network. So this type of network error correction codes is called subspace codes. Similarly, a metric was proposed to account for the discrepancy between the transmitted and received spaces and a coding theory based on this metric was developed. Silva, Koetter, and Kschischang \cite{Silva-K-K-rank-metric-codes} developed this approach further and explored the close relationship between subspace codes and rank-metric codes. For an overview of the development and some contributions in network error correction coding, refer to the survey papers \cite{KSK-subspace-codes, Zhang-survey-paper-NEC}.

In practical network communications, decoding only at sink nodes may not satisfy a plenty of applications. For example, many internal nodes of a network are also needed to decode the source messages. Thus, in \cite{Zhang-book} \cite{Yeung-book} (see also \cite{Tan-Yeung-Ho-Cai-Unified-Framework}), four useful classes of linear network codes were introduced, that are linear multicast, linear broadcast, linear dispersion, and generic network codes. These four classes of linear network codes possess properties of increasing strength. More generally, when channels of networks are noisy, besides the similar problems as mentioned above, we also have to face to the error correction problem. Hence, in order to take into account both information transmission and error correction simultaneously, it is necessary to consider error correction capability of four classes of linear network codes. Since decoding and error correction need much higher complexity than coding at internal nodes in general, in order to avoid long time delay, we consider decoding and coding separately and independently at internal nodes, that is, decoding problem of internal nodes is not under the consideration when we do encoding at internal nodes. Thus, we propose the concepts of linear network error correction (LNEC) multicast codes, LNEC broadcast codes, LNEC dispersion codes, and LNEC generic codes, respectively. Subsequently, for the four classes of LNEC codes, the extended Singleton bounds are proposed to characterize their error correction capabilities. Evidently, for the purpose of network error correction, it is expected to apply these codes meeting the corresponding extended Singleton bounds with equality, which, similarly, are also called maximum distance separable (MDS) codes. Motivated by it, the existence of the MDS codes is proved, which also shows that the proposed Singleton bounds are achievable. At last, to construct these classes of LNEC codes efficiently, several algorithms are obtained.

This paper is organized as follows. In the next section, we review linear network coding, four classes of linear network codes, and linear network error correction coding, particularly, the concepts and applications of the four important classes of linear network codes. Section \Rmnum{3} is devoted to LNEC multicast/broadcast/dispersion codes and \Rmnum{4} focuses on the same problems for LNEC generic codes. In Section \Rmnum{3}, because of the close relationships of the first three types of LNEC codes, we first introduce the definitions of LNEC multicast/broadcast/dispersion codes, and obtain the extended Singleton bounds for them. Further define LNEC multicast/broadcast/dispersion MDS codes, and prove the existence of these three classes of MDS codes by an algebra-based method. In Section \Rmnum{5}, we propose one class of algorithms for constructing these four classes of LNEC codes. The last section is the discussion in which we summarize the works done in this paper and indicate some topics for future research.

\section{Preliminaries}\label{Sec_preliminaries}

A communication network is represented by a finite acyclic directed graph $G=(V,E)$, where $V$ and $E$ are the sets of nodes and channels of the network, respectively. A direct edge $e=(i,j)\in E$ stands for a channel leading from node $i$ to node $j$. Node $i$ is called the tail of $e$ and node $j$ is called the head of $e$, denoted by $tail(e)$ and $head(e)$, respectively. Correspondingly, the channel $e$ is called an outgoing channel of $i$ or an incoming channel of $j$. For a node $i$, define $Out(i)$ as the set of outgoing channels of $i$ and $In(i)$ as the set of incoming channels of $i$. Mathematically,
\begin{align*}
Out(i)=\{e\in E:\ tail(e)=i\}, \quad \mbox{and} \quad In(i)=\{e \in E:\ head(e)=i\}.
\end{align*}
In a communication network, if a sequence of channels $(e_1,e_2,\cdots,e_m)$ satisfies $tail(e_1)=i,\ head(e_m)=j$, and $tail(e_{k+1})=head(e_k)$ for all $k=1,2,\cdots,m-1$, we say that the sequence $(e_1,e_2,\cdots,e_m)$ is a path from the node $i$ to the node $j$, or from the channel $e_1$ to the node $j$, or from the node $i$ to the channel $e_m$. For each channel $e\in E$, there exists a positive number $R_e$, said the capacity of $e$. The multiple channels between two nodes are allowed and thus assume reasonably that the capacity of any channel is 1 per unit time, that is, one field symbol can be transmitted over a channel in a unit time.

Throughout this paper, we consider single source networks, i.e., $|S|=1$, and the unique source node is denoted by $s$. A cut between the source node $s$ and a non-source node $t$ is a set of channels whose removal disconnects $s$ from $t$. For unit capacity channels, the capacity of a cut between $s$ and $t$ can be regarded as the number of channels in it, and the minimum of all capacities of cuts between the source node $s$ and a non-source node $t$ is called the minimum cut capacity between $s$ and $t$. A cut between $s$ and $t$ is called a minimum cut if its capacity achieves the minimum cut capacity between them.
Similarly, the concepts can be generalized to a collection $T$ of non-source nodes. A cut between the source node $s$ and $T$ is also defined as a set of channels whose removal disconnects $s$ from all $t\in T$. For unit capacity channels, the capacity of a cut between $s$ and $T$ can also be regarded as the number of channels in the cut, and the minimum of all capacities of the cuts between $s$ and $T$ is called the minimum cut capacity between them. A cut between $s$ and $T$ is called a minimum cut if its capacity achieves the minimum cut capacity between them.
In fact, if we expand the single source network $G=(V,E)$ into another $G_1=(V_1,E_1)$ by installing a new node $t_T$ which is connected from every node $t\in T$ by $C_t$ multiple unit capacity channels, where $C_t$ is the minimum cut capacity between $s$ and $t$ in $G$, then the minimum cut capacity between the source node $s$ and $T$ in $G$ is equal to the minimum cut capacity between $s$ and $t_T$ in $G_1$. Note that there may exist several minimum cuts between $s$ and $t$ (resp. $T$), but the minimum cut capacity between them is determined. The similar concept of minimum cut capacity for a collection $\xi$ of channels can be obtained, too. Specifically, we amend the network $G=(V,E)$ to obtain a new network $G_2=(V_2,E_2)$ by installing a new node $n_e$ for each channel $e=(i,j)\in \xi$ and replacing the channel $e$ by two new unit capacity channels $e_1$ and $e_2$, where $e_1$ is from the node $i$ to the node $n_e$ and $e_2$ is from the node $n_e$ to the node $j$, i.e., $e_1=(i, n_e)$ and $e_2=(n_e, j)$. Let $T_{\xi}$ be the set of nodes $n_e$, $e\in \xi$. Then the minimum cut capacity between the source node $s$ and the collection $\xi$ of channels in $G$ is defined as the minimum cut capacity between the source node $s$ and the the collection $T_{\xi}$ of non-source nodes in $G_2$.

Following the direction of the channels in $G$, there is an upstream-to-downstream order (ancestral topological order) on the channels in $E$ which is consistent with the partial order of all channels. The coordinates of all vectors and rows/columns of all matrices in the present paper are indexed according to this upstream-to-downstream order. In addition, if $L$ is such a matrix whose column vectors are indexed by a collection $B\subseteq E$ of channels according to an upstream-to-downstream order, then we use some symbol with subscript $e$, $e\in B$, such as $l_e$, to denote the column vector indexed by the channel $e$, and the matrix $L$ is written as column-vector form $L=\Big[l_e:\ e\in B\Big]$. If $L$ is a matrix whose row vectors are indexed by this collection $B$ of channels, then we use some symbol with $e$ inside a pair of brackets, such as $l(e)$, to denote the row vector corresponding to $e$, and the matrix $L$ is written as row-vector form $L=\Big[ l(e):\ e\in B \Big]$.

\subsection{Linear Network Coding}

In linear network coding, the source node $s$ generates messages and transmits them over the network by a linear network code. The source node $s$ has no incoming channels, but we introduce imaginary incoming channels for $s$ and assume that these imaginary incoming channels provide the source messages to $s$.
In this paper, we always use $\w$ to denote the information rate, and use $C_t$ (resp. $C_T$, $C_\xi$) to denote the minimum cut capacity between the unique source node $s$ and a non-source node $t$ (respectively, a collection $T$ of non-source nodes, a collection $\xi$ of channels), and define $\dt_t=C_t-\w$ (resp. $\dt_T=C_T-\w$, $\dt_\xi=C_\xi-\w$) as the redundancy of $t$ (resp. $T$, $\xi$). Because the information rate is $\w$ symbols per unit time, $s$ has $\w$ imaginary incoming channels, denoted by $d_1',d_2',\cdots,d_\w'$ and let $In(s)=\{d_1',d_2',\cdots,d_\w'\}$. The source messages are $\w$ symbols $\bX=[X_1\ X_2\ \cdots\ X_\w]$ arranged in a row vector where each $X_i$ is an element of the base field $\mathcal{F}$, and assume that they are transmitted to $s$ through the $\w$ imaginary incoming channels. Without loss of generality, assume that the message transmitted on the $i$th imaginary channel is the $i$th source message. Further, there is an $|In(i)|\times|Out(i)|$ matrix $K_i=[k_{d,e}]_{d\in In(i),e\in Out(i)}$ at each internal node $i$, say the local encoding kernel at $i$, where $k_{d,e}\in \mF$ is called the local encoding coefficient for the adjacent pair $(d,e)$ of channels. We use $U_e$ to denote the message transmitted over the channel $e$. Hence, at the source node $s$, we have $U_{d_i'}=X_i$, $1\leq i \leq \w$. In general, the message $U_e$ transmitted over the channel $e\in E$ is calculated recursively by the formulae:
$$U_e=\sum_{d\in In(tail(e))}k_{d,e}U_d.$$
Furthermore, it is not difficult to see that $U_e$ actually is a linear combination of the $\w$ source symbols $X_i$, $1\leq i\leq \w$, that is, there is an $\w$-dimensional column vector $f_e$ over the base field $\mF$ such that $U_e=\bX \cdot f_e$ (see also \cite{Zhang-book} \cite{Yeung-book}). This column vector $f_e$ is called the global encoding kernel of the channel $e$ and it can be determined by the local encoding kernels as follows:
$$f_e=\sum_{d\in In(tail(e))}k_{d,e}f_d,$$
with boundary condition that the vectors $f_{d_i'}$, $1\leq i \leq \w$, form the standard basis of the vector space $\mF^\w$.

\subsection{Linear Multicast, Linear Broadcast, Linear Dispersion, and Generic Network Codes}

In this subsection, we first recall the concepts of linear multicast, linear broadcast, and linear dispersion. Before that, we give a notation. For any subset $B\subseteq In(s)\cup E$, let $\mL(B)=\langle \{ f_e: e\in B \} \rangle$, where, as convenience, we use $\langle L \rangle$ to represent the subspace spanned by vectors in any collection $L$ of vectors.

\begin{defn}[{\cite[Definition 2.9]{Zhang-book}, \cite[Definition 19.12]{Yeung-book}}]\label{def_M_B_D}
An $\w$-dimensional linear network code on an acyclic network $G=(V,E)$ qualifies as a linear multicast, linear broadcast, and linear dispersion, respectively, if the following hold:
\begin{enumerate}
  \item $\dim(\mL(In(t)))=\w$ for every non-source node $t\in V$ with $C_t\geq \w$;
  \item $\dim(\mL(In(t)))=\min\{ \w, C_t \}$ for every non-source node $t\in V$;
  \item $\dim(\mL(In(T)))=\min\{ \w, C_T \}$ for every collection $T$ of non-source nodes.
\end{enumerate}
\end{defn}

For a linear multicast, a node $t$ can decode the source message vector $\bX$ successfully if and only if $C_t\geq \w$. An evident application of an $\w$-dimensional linear multicast is for multicasting source messages at information rate $\w$ to all or some of those non-source nodes with $C_t\geq \w$. For a linear broadcast, similarly, a node $t$ can decode the source message vector $\bX$ successfully if and only if $C_t\geq \w$. But for a node $t$ with $C_t<\w$, the set of all received global kernels, $\{ f_e:\ e\in In(t)\}$, can span a vector space with dimension $C_t$. An application of linear broadcast is that when the source node transmits the source messages at several different rates, the non-source node can decode them successfully once the rate satisfies $\w\leq C_t$. Moreover, when new channels of a node $t$ with $C_t<\w$ is connected from the network, just need to establish global encoding kernels of these new channels without modifying the linear broadcast. For a linear dispersion, a collection $T$ of non-source nodes can decode the source message vector $\bX$ successfully if and only if $C_T\geq \w$. And if $C_T<\w$, the received global encoding kernels of the collection $T$, i.e., $\{f_e:\ e\in \cup_{t\in T}In(t)\}$, can span a vector space with dimension $C_T$. An application of linear dispersion is in a two-tier network system consisting of a backbone network and many local area networks (LANs), where each LAN is connected to one or more nodes on the backbone network. The source messages with rate $\w$ generated by the source node $s$, in the backbone network, are to be
transmitted to every user on the LANs. With a linear dispersion on the backbone network, every user on a LAN can receive the source messages so long as the LAN acquires through the backbone network an aggregated max-flow from $s$ at least equal to $\w$. Moreover, new LANs can be established under the same criterion without modifying the linear dispersion on the backbone network.

It is easy to see that every linear dispersion is a linear broadcast, and every linear broadcast is a linear multicast. But a linear multicast is not necessarily a linear broadcast, and a linear broadcast is not necessarily a linear dispersion (refer to \cite[Example 2.10]{Zhang-book} or \cite[Example 19.13]{Yeung-book}).

Moreover, in \cite{Zhang-book}, the authors proposed an algebra-based definition of generic network codes, and then in \cite{Tan-Yeung-Ho-Cai-Unified-Framework}, \cite{Yeung-relation-dispersion-generic} and \cite{Yeung-Framework}, a graph-theory-based definition of generic network codes was given. In the following, we show the definition of generic network codes as a graph-theoretic interpretation.

\begin{defn}\label{def_generic codes}
Let $\{f_e: e\in E\}$ constitute a global description of an $\w$-dimensional linear network code over an acyclic network $G=(V,E)$. This code is called a generic network code if the following holds:
for any nonempty collection $\xi\subseteq E$ of channels with $|\xi|=\min\{ \w,C_{\xi} \}$, the global encoding kernels $f_e$, $e\in \xi$, are linearly independent, where again $C_{\xi}$ is the minimum cut capacity between $s$ and $\xi$.
\end{defn}

An equivalent condition of that one in Definition \ref{def_generic codes} will be given as follows.

\begin{lemma}
For an $\w$-dimensional linear network code on an acyclic network $G=(V,E)$, the following two conditions are equivalent:
\begin{enumerate}
  \item For any nonempty collection $\xi\subseteq E$ of channels, if $|\xi|=\min\{ \w,C_{\xi} \}$, then $f_e$, $e\in\xi$, are linearly independent.
  \item For any nonempty collection $\xi'\subseteq E$ of channels, $\dim(\langle\{ f_e: e\in \xi' \}\rangle)=\min\{\w,C_{\xi'}\}$.
\end{enumerate}
\end{lemma}
\begin{IEEEproof}
First, we prove $2)\Rightarrow 1)$. Let $\xi$ be an arbitrary collection of channels with $|\xi|=\min\{\w,C_{\xi}\}$ and satisfy the condition 2), i.e.,
$$\dim(\langle\{ f_e: e\in \xi \}\rangle)=\min\{\w,C_{\xi}\}=|\xi|.$$
This implies that $f_e$, $e\in\xi$, are linearly independent.

Next, we consider $1) \Rightarrow 2)$. Let $\xi'$ be an arbitrary nonempty collection of channels.

{\bf \textit{Case 1:}} If $|\xi'|=\min\{\w, C_{\xi'} \}$, then condition 1) implies that $f_e$, $e\in \xi'$, are linearly independent. That is,
$$\dim(\langle\{ f_e: e\in \xi' \}\rangle)=\min\{\w,C_{\xi'}\}.$$

{\bf \textit{Case 2:}} Otherwise $|\xi'|>\min\{ \w, C_{\xi'} \}$. There must exist a collection  $\xi \subseteq E$ of channels satisfying $\xi \subseteq \xi'$ and
$$C_{\xi}=|\xi|=\min\{ \w, C_{\xi'} \}.$$
For instance, setting $l=\min\{ \w,C_{\xi'} \}$, there are $l$ channel-disjoint paths from $s$ to $\xi'$. Let $\xi$ be the set of the last channels of all $l$ paths, and obviously, $\xi$ satisfies $\xi \subseteq \xi'$ and
$$C_{\xi}=|\xi|=\min\{ \w, C_{\xi'} \}.$$

By condition 1), we know that $f_e$, $e\in\xi$, are linearly independent, which leads to
$$\dim(\langle\{ f_e: e\in \xi \}\rangle)=|\xi|=\min\{\w,C_{\xi'}\}.$$
So we have
$$\dim(\langle\{ f_e: e\in \xi' \}\rangle)\geq \dim(\langle\{ f_e: e\in \xi \}\rangle)=\min\{\w,C_{\xi'}\}.$$
On the other hand, it is apparent that
$$\dim(\langle\{ f_e: e\in \xi' \}\rangle)\leq \min\{ \w, C_{\xi'} \}.$$
Combining the above, the lemma is proved.
\end{IEEEproof}

By this lemma, we can give a new equivalent definition of generic network codes, and this definition is more convenient for the following discussion and unified with the above three concepts.

\begin{defn}
An $\w$-dimensional linear network code over an acyclic network $G$ is called generic, if the following condition holds for any nonempty collection $\xi\subseteq E$ of channels:
$$\dim(\langle\{ f_e: e\in \xi \}\rangle)=\min\{ \w, C_{\xi} \}.$$
\end{defn}

Actually, for a generic network code, if a set of global encoding kernels can possibly be linearly independent, then it is linearly independent. In addition, every generic network code is a linear dispersion, but a linear dispersion is not necessarily a generic network code (refer to \cite[Example 2.15]{Zhang-book} or \cite[Example 19.31]{Yeung-book}).


\subsection{Linear Network Error Correction Coding}

For linear network error correction coding, we follow \cite{zhang-correction, Guang-MDS} in its notation and terminology. In the case that an error occurs on a channel $e$, the output of the channel is $\tilde{U}_e=U_e+Z_e$, where $U_e$ is the message that should be transmitted over the channel $e$ and $Z_e\in \mF$ is the error occurred on $e$. We also treat the error $Z_e$ as a message called \textit{error message}. Further, let error vector be an $|E|$-dimensional row vector $\bZ=[Z_e:\ e\in E]$ over the field $\mF$ with each component $Z_e$ representing the error occurred on the corresponding channel $e$. First, we introduce the extended network as follows. In network $G=(V,E)$, for each channel $e\in E$, we introduce an imaginary channel $e'$, which is connected to the tail of $e$ in order to provide the error message $Z_e$. The new network $\tilde{G}=(\tilde{V},\tilde{E})$ with imaginary channels is called the extended network of $G$, where $\tilde{V}=V$, $\tilde{E}=E\cup E'\cup \{d_1',d_2',\cdots, d_\w'\}$ with $E'=\{e': e\in E\}$. Clearly, $|E'|=|E|$. Then a linear network code on the original network $G$ can be amended to a linear network code on the extended network $\tilde{G}$ by setting $k_{e',e}=1$ and $k_{e',d}=0$ for all $d\in E\backslash\{e\}$. Note that, for each non-source node $i$ in $\tilde{G}$, $In(i)$ only includes the real incoming channels of $i$, that is, the imaginary channels $e'$ corresponding to $e\in Out(i)$ are not in $In(i)$. But for the source node $s$, we still define $In(s)=\{d_1',d_2',\cdots,d_\w'\}$. In order to distinguish two different types of imaginary channels, we say $d_i'$, $1\leq i\leq \w$, the \textit{imaginary message channels} and $e'$ for $e\in E$ the \textit{imaginary error channels}. Like linear network codes, we also define global encoding kernels $\f_e$ for all $e\in \tilde{E}$ in the extended network $\tilde{G}$, which is an $(\w+|E|)$-dimensional $\mF$-valued column vector and the entries can be indexed by all elements of $In(s)\cup E$. And further for imaginary message channels $d_i'$, $1\leq i \leq \w$, and imaginary error channels $e'\in E'$, set $\f_{d_i'}=1_{d_i'}$, and $\f_{e'}=1_e$, where $1_d$ is an $(\w+|E|)$-dimensional column vector which is the indicator function of $d\in In(s)\cup E$. And for other global encoding kernels $\f_e, e\in E$, we have recursive formulae:
$$\f_e=\sum_{d\in In(tail(e))}k_{d,e}\f_d+1_e.$$
We say $\f_e$ the extended global encoding kernel of the channel $e$ for the original network.

In ordinary LNEC coding, just error correction and decoding problems at sink nodes, which have no outgoing channels, are considered. But in practice, many internal nodes in the network are also required to decode the source messages. For ordinary LNEC codes, at each sink node $t$, the received message vector $\bU_t\triangleq \begin{bmatrix}\ti{U}_e:& e\in In(t)\end{bmatrix}$ and the decoding matrix $\F_t\triangleq \begin{bmatrix}\f_e:\ e\in In(t)\end{bmatrix}$ are available, and we have the following decoding equation:
\begin{align}\label{dec_equ}
\bU_t=(\bX\ \bZ)\F_t,
\end{align}
which can be used for decoding and error correction (refer to \cite{zhang-correction}\cite{Guang-MDS}). In our discussion, for each non-source node $t\in V$, we similarly define the received message vector $\bU_t$, and the decoding matrix $\F_t$, and still have decoding equation (\ref{dec_equ}). Moreover, sd mentioned above, in order to avoid long delay, both the decoding and error correction problem and the coding problem at internal nodes are independently, even if an internal node can decode the source messages successfully.

Similar to linear network codes \cite{Zhang-book}\cite{Yeung-book}, we can also define a linear network error correction code by either a local description or a global description.
\begin{defn}\
\begin{description}
  \item [{\bf Local Description of A Linear Network Error Correction Code.}]\

   An $\w$-dimensional $\mF$-valued linear network error correction code consists of all local encoding kernels at all internal nodes (including the source node $s$), i.e., $$ K=[k_{d,e}]_{d\in In(i), e\in Out(i)},$$
   that is an $|In(i)|\times |Out(i)|$ matrix for the node $i$, where $k_{d,e}\in \mF$ is the local encoding coefficient for the adjacent pair $(d,e)$ of channels with $d\in In(i)$, $e\in Out(i)$.
  \item[{\bf Global Description of A Linear Network Error Correction Code.}]\

  An $\w$-dimensional $\mF$-valued linear network error correction code consists of all extended global encoding kernels for all channels including imaginary message channels and imaginary error channels, which satisfy:
      \begin{enumerate}
        \item $\f_{d_i'}=1_{d_i'},\ 1 \leq i \leq \w$, and $\f_{e'}=1_e$, $e'\in E'$, where $1_d$ is an $(\w+|E|)$-dimensional column vector which is the indicator function of $d\in In(s) \cup E$;
        \item for other channel $e\in E$,
        \begin{align}\label{equ_ext_f}
        \f_e=\sum_{d\in In(tail(e))}k_{d,e}\f_d+1_e,
        \end{align}
        where $k_{d,e}\in \mF$ is the local encoding coefficient for the adjacent channel pair $(d,e)$ with $d\in In(tail(e))$, and again $1_e$ is an $(\w+|E|)$-dimensional column vector which is the indicator function of channel $e\in E$.
      \end{enumerate}
\end{description}
\end{defn}

Further, we give the following notation and definitions.

\begin{defn}
For each channel $e\in E$, the extended global encoding kernel $\f_e$ is written as follows:
\begin{equation*}
\f_e=\begin{bmatrix}
f_e(d_1')\\
\vdots\\
f_e(d_\w')\\
f_e(e_1)\\
\vdots\\
f_e(e_{|E|})\\
\end{bmatrix}
=\begin{bmatrix}
f_e\\
g_e\\
\end{bmatrix}
\end{equation*}
where
$f_e=\begin{bmatrix}
f_e(d_1')\\
\vdots\\
f_e(d_\w')
\end{bmatrix}
$ is an $\w$-dimensional column vector, and
$g_e=\begin{bmatrix}
f_e(e_1)\\
\vdots\\
f_e(e_{|E|})\\
\end{bmatrix}$
is an $|E|$-dimensional column vector.
\end{defn}

Recall that $\F_t=\begin{bmatrix}\f_e:\ e\in In(t)\end{bmatrix}$ is the decoding matrix of a non-source node $t\in V$. Denote by $\rt(d)$ the row vector of the decoding matrix $\F_t$ indicated by the channel $d\in In(s)\cup E$. These row vectors are of dimension $|In(t)|$. Hence,
\begin{equation*}
\tilde{F}_t=\begin{bmatrix}
\rt(d_1')\\
\vdots\\
\rt(d_\w')\\
\rt(e_1)\\
\vdots\\
\rt(e_{|E|})
\end{bmatrix}
=
\begin{bmatrix}
F_t\\
G_t
\end{bmatrix}
\end{equation*}
where $F_t=\begin{bmatrix}
\rt(d_1')\\
\vdots\\
\rt(d_\w')\end{bmatrix}$ and $G_t=\begin{bmatrix}\rt(e_1)\\ \vdots \\ \rt(e_{|E|}) \end{bmatrix}$ are two matrices of sizes $\w\times |In(t)|$ and $|E|\times |In(t)|$, respectively.

We use $\p$ to denote an error pattern which can be regarded as a set of channels. We call that an error message vector $\bZ$ matches an error pattern $\p$, if $Z_e=0$ for all $e\in E\backslash \p$. For any subset $B\subseteq In(s)\cup E$, define $\f_{e}^{(B)}=\begin{bmatrix} \f_e(d): d\in B \end{bmatrix}$, a $|B|$-dimensional column vector obtained from $\f_e$ by removing all entries $\f_e(d)$, $d\notin B$. Particularly, we give the following definition.

\begin{defn}[{\cite[Definition 2]{Guang-MDS}}]
For any error pattern $\p$ and any extended global encoding kernel $\f_e, e\in E$,
\begin{itemize}
  \item $\f_e^{\p}$ is an $(\w+|\p|)$-dimensional column vector obtained from $\f_e=[\f_e(d):\ d\in In(s)\cup E]$ by removing all entries $\f_e(d),d\notin In(s)\cup \p$.
  \item $f_e^{\p}$ is an $(\w+|E|)$-dimensional column vector obtained from $\f_e=[\f_e(d):\ d\in In(s)\cup E]$ by replacing all entries $\f_e(d)$, $d\notin In(s)\cup \p$, by $0$.
  \item $f_e^{\p^c}$ is an $(\w+|E|)$-dimensional column vector obtained from $\f_e=[\f_e(d):\ d\in In(s)\cup E]$ by replacing all entries $\f_e(d)$, $d\in In(s)\cup \p$, by $0$.
\end{itemize}
\end{defn}
Note that $f_e^{\p}+f_e^{\p^c}=\f_e$.

\section{Linear Network Error Correction Multicast/Broadcast/Dispersion Codes}

In this section, we discuss the first three types of LNEC codes, i.e., LNEC multicast/broadcast/dispersion codes.

\subsection{Notation and Definitions}

In order to solve the proposed problems, we introduce several new concepts most of which can be regarded as generalizations of the corresponding ones in ordinary linear network error correction codes. First, let $G=(V, E)$ be a single source acyclic network and $\{\f_e: e\in E \}$ constitute a global description of a linear network error correction code on $G$. Further, let $T$ be an arbitrary collection of non-source nodes and use $In(T)$ to denote $\cup_{t\in T}In(t)$.

\begin{defn}
The decoding matrix $\ti{F}_T$ for the collection $T$ is defined as:
$$\ti{F}_T=\begin{bmatrix}\f_e:& e\in In(T) \end{bmatrix}=\begin{bmatrix}\f_e: & e\in \cup_{t\in T}In(t) \end{bmatrix}.$$
Further, denote by $\row_T(d)$ the row vector of $\ti{F}_T$ indicated by the channel $d\in In(s)\cup E$. Then the decoding matrix $\ti{F}_T$ can be written as
$
\ti{F}_T=
\begin{bmatrix}
F_T\\
G_T\\
\end{bmatrix}
$,  where the matrix $F_T=\begin{bmatrix}
\row_T(d_i'):& 1\leq i\leq \w\end{bmatrix}$ of size $\w\times |In(T)|$ and the matrix $G_T=\begin{bmatrix}
\row_T(e): & e\in E
\end{bmatrix}$ of size $|E|\times |In(T)|$ .
\end{defn}

For the collection $T$ of non-source nodes, the following vector spaces are of importance.

\begin{defn}
Let $\p\subseteq E$ be an arbitrary error pattern. Define the following two vector spaces:
\begin{align*}
\Phi(T,G)=&\langle\{ \row_T(d_i'):\ 1\leq i\leq \w \}\rangle\\
       =&\{ (\bX\ \bzero)\cdot\ti{F}_T:\ \mbox{all $\w$-dimensional row vectors $\ul{{\bf X}}\in \mF^{\w}$} \},
\end{align*}
and
\begin{align*}
\Delta(T,\p,G)=&\langle\{ \row_T(e):\ e\in \p \}\rangle\\
=&\{ (\bzero\ \bZ)\cdot\ti{F}_T:\ \mbox{all $\ul{{\bf Z}}\in \mF^{|E|}$ matching error pattern $\p$} \},
\end{align*}
which is called the message space of $T$, and the error space of the error pattern $\p$ with respect to $T$, respectively.
\end{defn}

In some notation, when there is no ambiguity, $G$ is omitted usually and not omitted if necessary.

\begin{defn}
We say that an error pattern $\p_1$ is dominated by another error pattern $\p_2$ with respect to a collection $T$ of non-source nodes, if $\Delta(T,\p_1)\subseteq \Delta(T,\p_2)$ for any linear network error correction code. This relation is denoted by $\p_1 \prec_T \p_2$.
\end{defn}

\begin{defn}
The rank of an error pattern $\p$ with respect to a collection $T$ of non-source nodes is defined as
$$rank_T(\p)=\min\{ |\p'|:\ \p\prec_T \p'\},$$
where $|\p'|$ denotes the cardinality of the error pattern $\p'$.
\end{defn}

The above definition on the rank of an error pattern is abstract, and so in order to understand this concept more intuitively, we give the following lemma.

\begin{lemma}\label{lem_rank_T}
For an error pattern $\p$, introduce a source node $s_\p$. Let $\p=\{e_1,e_2,\cdots,e_l\}$, where $e_j\in In(i_j)$ for $1 \leq j \leq l$, and define $l$ new edges $e_j'=(s_\p, i_j)$. Replace each $e_j$ by $e_j'$ on the network $G$, that is, add $e_1',e_2',\cdots,e_l'$ on the network and delete $e_1,e_2,\cdots,e_l$ from the network. Then the rank of the error pattern $\p$ with respect to the collection $T$ in the original network is equal to the minimum cut capacity between $s_\p$ and $T$.
\end{lemma}

The proof of this lemma is similar to that of \cite[Lemma 1]{zhang-correction}, and, therefore, omitted. The readers are referred to that paper for technical details.

Review the concepts of the linear multicast/broadcast/dispersion in Definition \ref{def_M_B_D}. In the following, we refine their essential properties and apply them into LNEC coding. So we give the following definition which consists with the previous reference \cite{zhang-correction, Guang-MDS}, and actually is a restatement of the three properties for LNEC codes.

\begin{defn}\label{def_regular}\
\begin{enumerate}
  \item An $\w$-dimensional linear network error correction code is called {\it regular}, if $\dim(\Phi(t))=\w$ for any non-source node $t\in V$ with $C_t\geq \w$.
  \item An $\w$-dimensional linear network error correction code is called {\it strongly regular}, if $\dim(\Phi(t))=\min\{\w, C_t\}$ for any non-source node $t\in V$.
  \item An $\w$-dimensional linear network error correction code is called {\it strongly sup-regular}, if $\dim(\Phi(T))=\min\{\w, C_T\}$ for any collection $T$ of non-source nodes.
\end{enumerate}
\end{defn}

If a LNEC code has one of the three properties above, we say this LNEC code is the corresponding LNEC code. To be specific, an $\w$-dimensional LNEC code is said to be a \textit{LNEC multicast code}, if it is regular; an $\w$-dimensional LNEC code is said to be a \textit{LNEC broadcast code}, if it is strongly regular; and an $\w$-dimensional LNEC code is said to be a \textit{LNEC dispersion code}, it is is strongly sup-regular. Moreover, like the relationships of linear multicast, linear broadcast, and linear dispersion, every LNEC dispersion code is a LNEC broadcast code, and every LNEC broadcast codes is a LNEC multicast code.

Note that the above concepts only concern the aspect of information transmission. When another aspect of error correction is considered simultaneously, the following minimum distances paly an important role.

\begin{defn}\
\begin{enumerate}
  \item The minimum distance of a {\rm(}strongly{\rm)} regular linear network error correction code on $G$ at any non-source node $t\in V$ is defined as:
      \begin{align*}
      d_{\min}^{(t)}(G)&=\min\{ rank_t(\p): \Delta(t,\p)\cap\Phi(t)\neq\{\bzero\} \}\\
                       &=\min\{ |\p|: \Delta(t,\p)\cap\Phi(t)\neq\{\bzero\} \}\\
                       &=\min\{ \dim(\Delta(t,\p)): \Delta(t,\p)\cap\Phi(t)\neq\{\bzero\} \}.
      \end{align*}
  \item The minimum distance of a strongly sup-regular linear network error correction code on $G$ at any collection $T$ of non-source nodes is defined as:
      \begin{align}
      d_{\min}^{(T)}(G)&=\min\{ rank_T(\p): \Delta(T,\p)\cap\Phi(T)\neq\{\bzero\} \}\label{d_rank}\\
                       &=\min\{ |\p|: \Delta(T,\p)\cap\Phi(T)\neq\{\bzero\} \}.\label{d_p}\\
                       &=\min\{ \dim(\Delta(T,\p)): \Delta(T,\p)\cap\Phi(T)\neq\{\bzero\} \}.\label{d_dim}.
      \end{align}
\end{enumerate}
\end{defn}

\begin{rem}\
\begin{itemize}
  \item It is evident that $(\ref{d_dim})\leq (\ref{d_rank})\leq (\ref{d_p})$ since $\dim(\Delta(T,\p))\leq rank_T(\p)\leq |\p|$, and for $(\ref{d_p})\leq (\ref{d_rank})\leq (\ref{d_dim})$, it follows from the proof of \cite[Proposition 2]{Guang-MDS}.
  \item Similar to ordinary linear network error correction codes \cite{zhang-correction} and \cite{Guang-MDS}, we can also apply the minimum distance decoding principle to decode the source messages for each non-source node or each collection of non-source nodes, and the above minimum distances also fully characterize the error-detecting and error-correcting capabilities for each non-source node $t$ and each collection $T$ of non-source nodes, respectively.
\end{itemize}
\end{rem}


\subsection{Singleton Bounds}

In the following, we will indicate upper bounds on these minimum distances, which are closely similar to the Singleton bound in ordinary LNEC codes \cite{zhang-correction, Yang-refined-Singleton, Guang-MDS}, and thus we say them the extended Singleton bound and the weakly extended Singleton bound, respectively.

\begin{thm}[Extended Singleton Bound]\label{thm_extended_Singleton}
For any strongly sup-regular linear network error correction code on an acyclic network $G=(V,E)$, let $d_{\min}^{(T)}(G)$ be the minimum distance at any collection $T$ of non-source nodes in $V$, and then
\begin{align}\label{Singleton_T}
d_{\min}^{(T)}(G)\leq
\begin{cases}
\delta_T+1 & \mbox{ if } C_T\geq \w,\\
1          & \mbox{ if } C_T < \w.
\end{cases}
\end{align}
\end{thm}

Theorem \ref{thm_extended_Singleton} implies the following corollary immediately, when $T$ only contains one non-source node.

\begin{cor}[Weakly Extended Singleton Bound]\label{cor_weakly_extended_Singleton}
For any strongly regular linear network error correction code on an acyclic network $G=(V,E)$, let $d_{\min}^{(t)}(G)$ be the minimum distance at any non-source node $t\in V$, and then
\begin{align}\label{Singleton_t}
d_{\min}^{(t)}(G)\leq
\begin{cases}
\de_t+1 & \mbox{ if } C_t\geq \w,\\
1          & \mbox{ if } C_t < \w.
\end{cases}
\end{align}
\end{cor}
\begin{IEEEproof}[Proof of Theorem \ref{thm_extended_Singleton}]
For the acyclic network $G=(V,E)$, let $T$ be an arbitrary collection of non-source nodes in $V$, and $\ti{F}_T$ be the corresponding decoding matrix at the collection $T$. Since the considered LNEC code is strongly sup-regular, it follows that
$\dim(\Phi(T))=\min\{\w, C_T\}$ from Definition \ref{def_regular}.

To complete the proof, we discuss two cases below.

{\bf \textit{Case 1:}} $C_T\geq \w$, and thus $\dim(\Phi(T))=\min\{ \w, C_T \}=\w$.

Let the set of channels $\{e_1,e_2,\cdots,e_{C_T}\}$ be an minimum cut between $s$ and $T$ with an upstream-to-downstream order $e_1\prec e_2\prec \cdots \prec e_{C_T}$. And choose an error pattern $\p=\{e_{\w},e_{\w+1},\cdots,e_{C_T}\}$. Next, we will show that $\Delta(T,\p)\cap\Phi(T)\neq \{\bzero\}$.

Let $\bX$ be a source message vector and $\bZ$ be an error message vector. For each channel $e\in E$, we know $(\bX\ \bZ)\cdot \ti{f}_e=\ti{U}_e$, where recall that $\ti{f}_e$ is the extended global encoding kernel of $e$ and $\ti{U}_e$ is the output of the channel $e$ .
Let $\ti{U}_{e_1}=\ti{U}_{e_2}=\cdots=\ti{U}_{e_{\w-1}}=0$. Since the rank of the matrix
$\begin{bmatrix}\ti{f}_{e_1}&\ti{f}_{e_2}&\cdots&\ti{f}_{e_{\w-1}} \end{bmatrix}$ is at most $(\w-1)$, it follows that there exists a nonzero message vector $\bX_1$ and an all-zero error message vector $\bZ_1=\bzero$ such that
\begin{align*}
&\begin{pmatrix}\bX_1 & \bZ_1\end{pmatrix}\cdot\begin{bmatrix}\ti{f}_{e_1}&\ti{f}_{e_2}&\cdots&\ti{f}_{e_{\w-1}} \end{bmatrix}\\
=&\bX_1\cdot\begin{bmatrix}f_{e_1}&f_{e_2}&\cdots&f_{e_{\w-1}}\end{bmatrix}\\
=&\begin{bmatrix}\ti{U}_{e_1}&\ti{U}_{e_2}&\cdots&\ti{U}_{e_{\w-1}} \end{bmatrix}=\bzero.
\end{align*}
Furthermore, since the LNEC code is strongly sup-regular, this implies
\begin{align*}
\begin{pmatrix}\bX_1 & \bZ_1 \end{pmatrix}\cdot\begin{bmatrix}\ti{f}_{e_1}&\ti{f}_{e_2}&\cdots&\ti{f}_{e_{C_T}}
\end{bmatrix}
=\begin{bmatrix}\ti{U}_{e_1}&\ti{U}_{e_2}&\cdots&\ti{U}_{e_{C_T}} \end{bmatrix}\neq \bzero.
\end{align*}
Assume the contrary, i.e., $[\tilde{U}_{e_1}\ \tilde{U}_{e_2}\ \cdots\ \tilde{U}_{e_{C_T}}]=\bzero$. Note that $\{e_1,e_2,\cdots,e_{C_T}\}$ is a minimum cut between $s$ and $T$ and $\bZ_1=\bzero$. It follows that
$$\bU_T\triangleq[\tilde{U}_e:\ e\in In(T)]=\bzero,$$
which implies that $(\bX_1\ \bzero)\tilde{F}_T=\bzero$ from the decoding equation $(\bX_1\ \bZ_1)\tilde{F}_T=\bU_T$. Therefore, we obtain $\bX_1=\bzero$ from $\dim(\Phi(T))=\Rank(F_T)=\w$ as the strongly sup-regular property of the LNEC code. This contradicts $\bX_1\neq\bzero$.

On the other hand, there exists another source message vector $\bX_2=\bzero$ and another error message vector $\bZ_2$ satisfying the conditions that $\bZ_2$ matches the error pattern $\p=\{e_{\w},e_{\w+1},\cdots,e_{C_T}\}$, and
$$
\begin{pmatrix}\bX_2&\bZ_2\end{pmatrix}\cdot\begin{bmatrix}\ti{f}_{e_1}&\cdots&\ti{f}_{e_{C_T}} \end{bmatrix}
=\begin{bmatrix}\ti{U}_{e_1}&\cdots&\ti{U}_{e_{C_T}} \end{bmatrix}.
$$
First, it is evident that $\bZ_2\neq \bzero$ because $[\tilde{U}_{e_1}\ \tilde{U}_{e_2}\ \cdots\ \tilde{U}_{e_{C_T}}]\neq \bzero$.
And, since $e_{\w}\prec e_{\w+1}\prec \cdots\prec e_{C_T}$, for any $e\in \p$, we can set sequentially:
$$Z_e=\ti{U}_e-\sum_{d\in In(tail(e))}k_{d,e}\ti{U}_d'$$
with the boundary condition that $Z_e=0$ for all $e\in E\backslash\p$,
where $\tilde{U}_d'$ is the output of the channel $d$ in this case.

Combining the above, we deduce
$$\begin{pmatrix}\bX_1&\bzero\end{pmatrix}\cdot\ti{F}_T
=\begin{pmatrix}\bzero&\bZ_2\end{pmatrix}\cdot\ti{F}_T,$$
which, together with the fact that $\bZ_2\neq \bzero$ matches the error pattern $\p$, proves that
$$\Phi(T)\cap\Delta(T,\p)\neq \{\bzero\}.$$
In other words, $d_{\min}^{(T)}(G)\leq \de_T+1$ for any collection $T$ of non-source nodes with $C_T\geq \w$.

{\bf \textit{Case 2:}} $C_T<\w$ and thus $\dim(\Phi(T))=\min\{\w, C_T\}=C_T$.

Similarly, still let $\{e_1,e_2,\cdots,e_{C_T}\}$ be an arbitrary minimum cut between $s$ and $T$ in an upstream-to-downstream order $e_1\prec e_2\prec \cdots \prec e_{C_T}$. Further, let $\ti{U}_{e_{1}}=\ti{U}_{e_2}=\cdots=\ti{U}_{e_{C_T-1}}=0$ and $\ti{U}_{e_{C_T}}=1$. Since the rank of the $\w\times C_T$ matrix $\begin{bmatrix}f_{e_{1}}&f_{e_{2}}&\cdots&f_{e_{C_T}}\end{bmatrix}$ is $C_T$, there must exist a nonzero message vector $\bX_1$ and an all-zero error message vector $\bZ_1=\bzero$ such that
\begin{align*}
&\begin{pmatrix}\bX_1&\bZ_1\end{pmatrix}\cdot\begin{bmatrix}\ti{f}_{e_1}&\ti{f}_{e_2}&\cdots&\ti{f}_{e_{C_T}}\end{bmatrix}\\
=&\bX_1\cdot\begin{bmatrix}f_{e_1}&f_{e_2}&\cdots&f_{e_{C_T}}\end{bmatrix}\\
=&\begin{bmatrix}\ti{U}_{e_{1}}& \ti{U}_{e_2} & \cdots & \ti{U}_{e_{C_T}}\end{bmatrix}\\
=&\begin{bmatrix}\bzero_{C_T-1}&1\end{bmatrix},
\end{align*}
where $\bzero_{C_T-1}$ represents a $(C_T-1)$-dimensional all-zero row vector. On the other hand, let $\bX_2=\bzero$ be another source message vector and $\bZ_2$ be another error message vector satisfying $Z_{e_{C_T}}=1$ and $Z_e=0$ for other channels
$e\in E\backslash \{e_{C_T}\}$. Then we easily have
\begin{align*}
&\begin{pmatrix}\bX_2&\bZ_2\end{pmatrix}\cdot\begin{bmatrix}\ti{f}_{e_1}&\ti{f}_{e_2}&\cdots&\ti{f}_{e_{C_T}} \end{bmatrix}\\
=&\begin{bmatrix}\ti{U}_{e_{1}}& \ti{U}_{e_2} & \cdots & \ti{U}_{e_{C_T}}\end{bmatrix}\\
=&\begin{bmatrix}\bzero_{C_T-1}&1\end{bmatrix}.
\end{align*}

Therefore, it follows that
$$\begin{pmatrix}\bX_1&\bzero\end{pmatrix}\cdot\ti{F}_T
=\begin{pmatrix}\bzero&\bZ_2\end{pmatrix}\cdot\ti{F}_T,$$
which, together with $\bZ_2$ matching the error pattern $\p=\{e_{C_T}\}$. This implies that
$\Phi(T)\cap \Delta(T,\p)\neq\{\bzero\}.$
That is, $d_{\min}^{(T)}(G)\leq 1$ for all collections $T$ of non-source nodes with $C_T<\w$.

Combining the two cases, the proof is completed.
\end{IEEEproof}

If the above (weakly) extended Singleton bound is achievable, we adopt the convention that the codes meeting Singleton bound with equality are called maximum distance separable (MDS) codes. Thus, we present the following definition.

\begin{defn}
An $\w$-dimensional LNEC code on a network $G$ is called LNEC multicast MDS code, LNEC broadcast MDS code, and LNEC dispersion MDS code, respectively, or multicast MDS code, broadcast MDS code, and dispersion MDS code for short, if the following hold respectively:
\begin{enumerate}
  \item this LENC code is regular and $d_{\min}^{(t)}(G)=\de_t+1$ for any non-source node $t\in V$ with $C_t\geq \w$;
  \item this LENC code is strongly regular and the weakly extended Singleton bound (\ref{Singleton_t}) is satisfied with equality for any non-source node $t\in V$;
  \item this LENC code is strongly sup-regular and the extended Singleton bound (\ref{Singleton_T}) is satisfied with equality for any nonempty collection $T$ of non-source nodes.
\end{enumerate}
\end{defn}

\begin{rem}\label{rem_relation-m-b-d}
It is not difficult to observe that every dispersion MDS code is a broadcast MDS code, and every broadcast MDS code is a multicast MDS code, but not vice versa.
\end{rem}

\subsection{The Existence of Linear Network Error Correction Multicast/Broadcast/Dispersion MDS Codes}

In this subsection, we will study the achievability of the given Singleton bounds above, in other words, we will consider the existence of LNEC multicast/broadcast/dispersion MDS codes. Before discussion further, we need some notation and lemmas as follows.

Again let $G$ be an acyclic network and $T$ be a collection of non-source nodes with $C_T\geq \w$. Define $R_T(\dt_T,G)$ as the set of error patterns $\p$ satisfying $|\p|=rank_T(\p)=\dt_T$, i.e.,
$$
R_T(\dt_T,G)=\{\mbox{ error pattern }\p:\ |\p|=rank_T(\p)=\dt_T \}.
$$
When $T$ contains only one non-source node $t$ with $C_t\geq \w$, $R_T(\dt_T,G)$ is written as:
$$
R_t(\dt_t,G)=\{\mbox{ error pattern }\p:\ |\p|=rank_t(\p)=\dt_t\}.
$$
When there is no ambiguity, $R_T(\dt_T,G)$ and $R_t(\dt_t,G)$ will be abbreviated as $R_T(\dt_T)$ and $R_t(\dt_t)$, respectively.

\begin{lemma}[{\cite[Corollary 4]{Guang-MDS}}]\label{lem_path}
For each $t\in V$ with $C_t\geq \w$ and any error pattern $\p\in R_t(\dt_t)$, there exist $(\w+\dt_t)$ channel-disjoint paths from either $In(s)=\{d_1',d_2',\cdots,d_\w'\}$ or $\p'=\{e': e\in \p\}$ to $t$, and the $(\w+\dt_t)$ paths satisfy the following properties:
\begin{enumerate}
  \item there are exactly $\dt_t$ paths from $\p'$ to $t$, and $\w$ paths from $In(s)$ to $t$;
  \item these $\dt_t$ paths from $\p'$ to $t$ start with the distinct channels in $\p'$ and for each path, if it starts with $e'\in \p'$, then it passes through $e\in \p$.
\end{enumerate}
\end{lemma}

\begin{lemma}[{\cite[Lemma 19.17]{Yeung-book} and \cite[Lemma 1]{Koetter-Medard-algebraic}}]\label{lem_poly}
Let $f(x_1,x_2,\cdots,x_n)$ be a nonzero polynomial with coefficients in a field $\mF$. If $|\mF|$ is greater than the degree of $f$ for any $x_i$, $1\leq i \leq n$, then there exist $a_1,a_2,\cdots,a_n\in \mF$ such that
$f(a_1,a_2,\cdots,a_n)\neq 0$.
\end{lemma}

The following theorems show the existence of LNEC multicast/broadcast/dispersion MDS codes.

\begin{thm}\label{thm_e_m}
Let $G=(V,E)$ be a single source acyclic network. There exists an $\w$-dimensional linear network error correction multicast MDS code on $G$, if the size of the base field satisfies:
$$|\mF|>\sum_{t\in V:\ C_t\geq \w}|R_t(\dt_t)|.$$
\end{thm}
\begin{IEEEproof}
Let $t$ be an arbitrary non-source node in $V$ with $C_t\geq \w$, and $\p$ be an arbitrary error pattern in $R_t(\de_t)$. Recall that $\ti{F}_t=\begin{bmatrix}\ti{f}_e:& e\in In(t)\end{bmatrix}$ is the decoding matrix at $t$, and $\ti{F}_t^{\p}=\begin{bmatrix}\ti{f}_e^{\p}:& e\in In(t)\end{bmatrix}$. It is not difficult to see that each entry of $\ti{F}_t$ (obviously, $\ti{F}_t^{\p}$) is a polynomial of local encoding coefficients $k_{d,e}$ for channel adjacent pairs $(d,e)$, $d,e\in In(s)\cup E$. Define an $(\w+\de_t)\times(\w+\de_t)$ matrix $A_t(\p)$ and an $|In(t)|\times(\w+\de_t)$ matrix $B_t(\p)$, where all entries of $A_t(\p)$ and $B_t(\p)$ are variables taking values in the base field $\mF$.

First, we indicate that the determinant $\det(A_t(\p)\ti{F}_t^{\p}B_t(\p))$ is a nonzero polynomial. Since $rank_t(\p)=\de_t$ and Lemma \ref{lem_path}, there exist $(\w+\de_t)$ channel-disjoint paths satisfying the following conditions:
\begin{enumerate}
  \item there are exactly $\de_t$ paths from $\p'$ to $t$, and $\w$ paths from $In(s)$ to $t$;
  \item these $\de_t$ paths from $\p'$ to $t$ start with the distinct channels in $\p'$ and for each path, if it starts with $e'\in \p'$, then it passes through $e\in \p$.
\end{enumerate}
Put $k_{d,e}=1$ for all adjacent pairs of channels $(d,e)$ along any one of the chosen $(\w+\de_t)$ channel-disjoint paths, and $k_{d,e}=0$, otherwise. This means that in this case $\ti{F}_t^{\p}$ contains an $(\w+\de_t)\times(\w+\de_t)$ identity submatrix. Thus, we can take the proper values in $\mF$ for the entries of $A_t(\p)$ and $B_t(\p)$ such that
$$A_t(\p)\ti{F}_t^{\p}B_t(\p)=I_{\w+\de_t},$$
that is, $\det(A_t(\p)\ti{F}_t^{\p}B_t(\p))=1,$ which shows that $\det(A_t(\p)\ti{F}_t^{\p}B_t(\p))$ is a nonzero polynomial.

In the following, we will show that the degree of each indeterminate $k_{d,e}$ in the nonzero polynomial $\det(A_t(\p)\ti{F}_t^{\p}B_t(\p))$ is $1$ at most.
Let the matrix $\ti{M}=\begin{bmatrix}\ti{f}_e:& e\in E\end{bmatrix}$, where put all the extended global encoding kernels in juxtaposition according to the given upstream-to-downstream order. Further let the matrix $\ti{A}=\left[\begin{smallmatrix} A \\ I\end{smallmatrix}\right]$, where $I$ represents an $|E|\times |E|$ identity matrix, and $A=(k_{d,e})_{d\in In(s), e\in E}$ is an $\w\times|E|$ matrix with $k_{d,e}=0$ for $e\notin Out(s)$ and $k_{d,e}$ being the local encoding coefficient for $e\in Out(s)$. And let the system transfer matrix $F=(k_{d,e})_{d\in E, e\in E}$ be an $|E|\times |E|$ matrix with $k_{d,e}$ being the local encoding coefficient for $head(d)=tail(e)$ and $k_{d,e}=0$ for $head(d)\neq tail(e)$. Therefore, we have the formula $$\ti{M}=\ti{A}(I-F)^{-1}.$$
This formula is similar to the Koetter-M\'{e}dard Formula\cite{Koetter-Medard-algebraic} in linear network coding, and first appeared in \cite{zhang-correction}. Furthermore, there exists an $(\w+|\p|)\times(\w+\mE)$ matrix $A_{\p}$ and an $|E| \times |In(t)|$ matrix $B_t$ such that
$$\ti{F}_t^{\p}=A_{\p}\ti{M}B_t=A_{\p}\ti{A}(I-F)^{-1}B_t.$$
Consequently,
$$A_t(\p)\ti{F}_t^{\p}B_t(\p)=A_t(\p)\cdot A_{\p}\cdot \ti{A}\cdot (I-F)^{-1}\cdot B_t\cdot B_t(\p).$$
Further, notice that
\begin{align}
&\det\Big( \begin{bmatrix}A_t(\p)A_{\p}\ti{A}&\bzero_{(\w+\de_t)\times(\w+\de_t)}\\I-F&B_tB_t(\p) \end{bmatrix} \Big)\nonumber\\
=&\det\Big( \begin{bmatrix}A_t(\p)A_{\p}\ti{A}&-A_t(\p)A_{\p}\ti{A}(I-F)^{-1}B_tB_t(\p)\\I-F&\bzero_{|E|\times (\w+\de_t)}\end{bmatrix} \Big)\nonumber\\
=&(-1)^*\det(I-F)\cdot\det(A_t(\p)A_{\p}\ti{A}(I-F)^{-1}B_tB_t(\p))\nonumber\\
=&\det(A_t(\p)A_{\p}\ti{A}(I-F)^{-1}B_tB_t(\p))\cdot(-1)^*\label{1}\\
=&\det(A_t(\p)\ti{F}_t^{\p}B_t(\p))\cdot(-1)^*,
\end{align}
where  $\bzero_{a\times b}$ represents an $a\times b$ all-zero matrix, and (\ref{1}) follows from $\det(I-F)=1$ as $F$ is an upper triangular matrix and all elements of main diagonal are zeros. This implies that the degree of each indeterminate $k_{d,e}$ in the polynomial $\det(A_t(\p)\ti{F}_t^{\p}B_t(\p))$ is $1$ at most.

As this conclusion can be applied to every non-source node $t\in V$ with $C_t\geq \w$ and every $\p\in R_t(\de_t)$, it follows that the polynomial
$$\prod_{t\in V:\ C_t\geq \w}\prod_{\p\in R_t(\de_t)}\det(A_t(\p)\ti{F}_t^{\p}B_t(\p))$$
is also a nonzero polynomial over the base field $\mF$, and the degree of each indeterminate $k_{d,e}$ is at most $\sum_{t\in V:\ C_t\geq \w}|R_t(\de_t)|$. Together with Lemma \ref{lem_poly}, this proves that if
$$|\mF|>\sum_{t\in V:\ C_t\geq \w}|R_t(\de_t)|,$$
we can take scalar values in $\mF$ for all indeterminates such that
$$\prod_{t\in V:\ C_t\geq \w}\prod_{\p\in R_t(\de_t)}\det(A_t(\p)\ti{F}_t^{\p}B_t(\p))\neq 0,$$
which further means all determinants are nonzero. This shows that for every non-source node $t$ with $C_t\geq \w$ and every error pattern $\p$ with $|\p|=rank_t(\p)=\de_t$, $\Rank(\ti{F}_t^{\p})=\w+\de_t$, or equivalently, $\Phi(t)\cap \Delta(t,\p)=\{\bzero\}$.

In addition, for any error pattern $\eta$ with $|\eta|<\de_t$, there exists an error pattern $\p\in R_t(\de_t)$ such that $\eta \prec_t\p$ from Lemma \ref{lem_rank_T}. Therefore, we obtain that
$d_{\min}^{(t)}(G)\geq \de_t+1$
for all non-source nodes $t\in V$ with $C_t\geq \w$. The proof is completed by combining with the weakly extended Singleton bound (see Corollary \ref{cor_weakly_extended_Singleton}).
\end{IEEEproof}

We have indicated the existence of LNEC multicast MDS codes, which actually can lead to the existence of LNEC broadcast/dispersion MDS codes. First, we show the existence of LNEC broadcast MDS codes.

\begin{thm}\label{thm_e_b}
Let $G=(V,E)$ be a single source acyclic network. There exists an $\w$-dimensional LNEC broadcast MDS code on $G$, if the size of the base field satisfies:
$$|\mF|>\sum_{t\in V:\ C_t\geq \w}|R_t(\dt_t)|+|V_2|,$$
where $V_2\subseteq V$ is the set of all non-source nodes $t\in V$ with $C_t<\w$.
\end{thm}
\begin{IEEEproof}
From the network $G$, we construct a new network $G'=(V',E')$ as follows:
\begin{enumerate}
  \item install a new node $t'$ for each non-source node $t$ with $C_t<\w$;
  \item install $C_t$ multiple channels from $t$ to $t'$ and $(\w-C_t)$ multiple channels from $s$ to $t'$.
\end{enumerate}
Therefore, for the new network $G'$, we obtain
\begin{align*}
V'&=V\cup\{ t':\ t\in V \mbox{ with } C_t<\w\}=V\cup\{t': t\in V_2 \},\\
E'&=E\cup\cup_{\mbox{\scriptsize all }t'}In(t')=E\cup\cup_{t': t\in V_2}In(t'),
\end{align*}
and obviously, $|V'|=|V|+|V_2|$ and $|E'|=|E|+\w|V_2|$.

Now from Theorem \ref{thm_e_m} we know that there exists an $\w$-dimensional multicast MDS code on $G'$ for some finite filed $\mF$. And further let $\{ \ti{f}_e:\ e\in E' \}$ constitute a global description of this multicast MDS code. Actually, it is not difficult to check that $\{\f_e^{(In(s)\cup E)}:\ e\in E\}$ constitutes a global description of a LNEC code on the original network $G$. To be specific, for any $e\in E$, the following equality holds:
$$\f_e^{(In(s)\cup E)}=\sum_{d\in In(tail(e))}k_{d,e}\f_d^{(In(s)\cup E)}+1_e^{(In(s)\cup E)},$$
and the boundary conditions below are also satisfied:
$$\f_{d_i'}^{(In(s)\cup E)}=1_{d_i'}^{(In(s)\cup E)},\quad 1\leq i \leq \w,$$
where $1_d$ is an $(\w+|E'|)$-dimensional column vector which is the indicator function of $d\in In(s)\cup E'$.
Next, we will prove that $\{\f_e^{(In(s)\cup E)}: e\in E\}$ actually constitutes an $\w$-dimensional $\mF$-valued broadcast MDS code on $G$.

{\bf \textit{Case 1:}} For any non-source node  $t\in V$ with $C_t<\w$, since
$$\Phi(t,G)=\Phi(t,G')=\langle \{ \row_t(d_i'):\ 1\leq i \leq \w \} \rangle,$$
we deduce that
\begin{align*}
\dim(\Phi(t,G))=\dim(\Phi(t,G'))=\dim(\langle \{ f_e: e\in In(t) \} \rangle) \geq \dim(\langle \{ f_e: e\in Out(t)\cap In(t') \} \rangle)=C_t,
\end{align*}
where the inequality follows as each $f_e$, $e\in Out(t)\cap In(t')$, is a linear combination of all $f_e$, $e\in In(t)$, and the last equality follows because the considered LNEC code on $G'$ is regular.
And certainly, $\dim(\Phi(t,G'))\leq C_t$. So it is shown that
$$\dim(\Phi(t,G))=\dim(\Phi(t,G'))=C_t.$$

In addition, the weakly extended Singleton bound indicated $d_{\min}^{(t)}(G)\leq 1$, and evidently, $d_{\min}^{(t)}(G)>0$. Thus one has $d_{\min}^{(t)}(G)=1$ for all $t\in V$ with $C_t<\w$.

Combining the above, it follows that for any non-source node $t\in V$ with $C_t<\w$, $\dim(\Phi(t,G))=\min\{\w, C_t\}=C_t$ and $d_{\min}^{(t)}(G)=1$.

{\bf \textit{Case 2:}} For any non-source node $t\in V$ with $C_t\geq \w$, similarly,
$$\Phi(t,G)=\Phi(t,G')=\langle \{ \row_t(d_i'):\ 1\leq i \leq \w \} \rangle,$$
which, together with the fact that the considered LNEC code on $G'$ is multicast MDS, implies
\begin{align*}
\dim(\Phi(t,G))=\dim(\Phi(t,G'))=\min\{\w, C_t\}=\w,
\end{align*}
and
\begin{align*}
d_{\min}^{(t)}(G')=\min\{ |\p|: \Delta(t,\p,G')\cap\Phi(t,G')\neq\{\bzero\} \}=\de_t+1.
\end{align*}

Let $\p\subseteq E'$ be an arbitrary error pattern satisfying
$|\p|=\de_t+1$ and $\Delta(t,\p,G')\cap\Phi(t,G')\neq\{\bzero\}$. We claim that $\{\row_t(e): e\in \p\}$ are linearly independent. Conversely, there exists $\p_1\subsetneq \p$ such that $|\p_1|<\de_t+1$ and
$$\Delta(t,\p_1,G')\cap \Phi(t,G')=\Delta(t,\p,G')\cap \Phi(t,G')\neq \{\bzero\},$$
which violates the condition $d_{\min}^{(t)}(G')=\de_t+1$. Further, note the fact $\row_t(e)=\bzero$ for all channels $e\in E'\backslash E$ as no path exists from $e$, $e\in E'\backslash E$, to any channel $d$, $d\in In(t)$. This implies that $\p\subseteq E$. Therefore, it follows $\Delta(t,\p,G')=\Delta(t,\p,G)$, which, together with $\Phi(t,G')=\Phi(t,G)$ and
$$\Delta(t,\p,G)\cap\Phi(t,G)=\Delta(t,\p,G')\cap\Phi(t,G')\neq \{\bzero\},$$
leads to
$$d_{\min}^{(t)}(G)=\min\{ |\p|: \Delta(t,\p,G)\cap\Phi(t,G)\neq \{\bzero\} \}=\de_t+1.$$
Therefore, for any non-source node $t\in V$ with $C_t\geq \w$, one obtains $\dim(\Phi(t,G))=\w$ and $d_{\min}^{(t)}(G)=\de_t+1$.

At last, we take the field size into account. By Theorem \ref{thm_e_m}, we know that if
$$|\mF|>\sum_{t\in V':\ C_t\geq \w}|R_t(\de_t,G')|,$$
there exists an $\w$-dimensional $\mF$-valued multicast MDS code on $G'$, and further there exists an $\w$-dimensional $\mF$-valued broadcast MDS code on $G$. And one has
\begin{align*}
\sum_{t\in V': C_t\geq \w}|R_t(\de_t,G')|=&\sum_{t\in V: C_t\geq \w}|R_t(\de_t,G)|+\sum_{t\in V_2}|R_{t'}(\de_{t'},G')|\\
=&\sum_{t\in V: C_t\geq \w}|R_t(\de_t,G)|+\sum_{t\in V_2}|R_{t'}(0,G')|\\
=&\sum_{t\in V: C_t\geq \w}|R_t(\de_t,G)|+|V_2|.
\end{align*}
The proof is accomplished.
\end{IEEEproof}

Next, we show the existence of LNEC dispersion MDS codes.

\begin{thm}\label{thm_e_d}
Let $G=(V,E)$ be a single source acyclic network. There exists an $\w$-dimensional $\mF$-valued LNEC dispersion MDS code on $G$, if the size of the base field $\mF$ satisfies:
$$|\mF|>\sum_{T\in \mT: C_T\geq \w}|R_{t_T}(\dt_T, G')|+|V_3|,$$
where $\mT$ is the set of all collections of non-source nodes, $V_3$ is the set of all collections $T\in \mT$ with $C_T<\w$, and $G'$ and $t_T$ are a new network and a new node corresponding $G$ and $T\in \mT$ with $C_T\geq \w$ respectively, as described at the beginning of the proof.
\end{thm}
\begin{IEEEproof}
From the network $G$, we construct a new network as follows:
\begin{enumerate}
  \item for any $T\in \mT$, install a new node $t_T$;
  \item install $C_t$ multiple channels from $t$ to $t_T$ for all $t\in T$.
\end{enumerate}
This new network is denoted by $G'=(V',E')$, where
$V'=V\cup\{ t_T: T\in \mT \}$ and $E'=E\cup\cup_{T\in \mT} In(t_T)$. And we have
\begin{align*}
|V'|&=|V|+\sum_{i=1}^{|V|-1}{|V|-1 \choose i}=|V|+(2^{|V|-1}-1),\\
|E'|&=|E|+\sum_{T\in \mT}|In(t_T)|=|E|+\sum_{T\in \mT}\sum_{t\in T}C_t.
\end{align*}

By Theorem \ref{thm_e_b}, there exists an $\w$-dimensional broadcast MDS code on $G'$ for some filed $\mF$ , and let $\{ \ti{f}_e: e\in E' \}$ constitute a global description of an $\w$-dimensional LNEC broadcast MDS code on $G'$. Actually, $\{\f_e^{(In(s)\cup E)}: e\in E\}$ constitutes a global description of a LNEC code on $G$. In the following, we will present that $\{\f_e^{(In(s)\cup E)}: e\in E\}$ also constitutes an $\w$-dimensional $\mF$-valued LNEC dispersion MDS code on $G$.

We first indicate that this LNEC code on $G$ is strongly sup-regular. For any $T\in \mT$, since
$$\Phi(T,G)=\Phi(T,G')=\langle \{ \row_T(d_i'): 1\leq i \leq \w \}\rangle $$
and the original LNEC code on $G'$ is strongly regular, it follows that
\begin{align*}
\dim(\Phi(T,G))=&\dim(\Phi(T,G'))\\
               =&\dim(\langle \{ f_e:\ e\in In(T) \} \rangle)\\
               \geq& \dim(\langle \{ f_e:\ e\in In(t_T) \} \rangle)\\
               =&\min\{\w,C_{t_T}\}=\min\{\w,C_{T}\},
\end{align*}
where again the above inequality follows as each $f_e$, $e\in In(t_T)$, is a linear combination of the vectors in $\{ f_e: e\in In(T) \}$, and the last step holds because of $C_{t_T}=C_{T}$.
On the other hand, $\dim(\Phi(T,G'))\leq \min\{\w,C_{T}\}$. Thus,
$$\dim(\Phi(T,G))=\dim(\Phi(T,G'))=\min\{\w,C_{T}\}.$$

Next, we prove that the inequality (\ref{Singleton_T}) in extended Singleton bound holds with equality for this LNEC code. Since $\{ \ti{f}_e: e\in E' \}$ constitutes a global description of an $\w$-dimensional $\mF$-valued LNEC broadcast MDS code on $G'$, one has
$$d_{\min}^{(t_T)}(G')=\begin{cases}\de_{t_T}+1=\de_T+1 & \mbox{ if } C_{t_T}\geq \w\mbox{ i.e., }C_T\geq \w,\\1& \mbox{ if } C_{t_T}<\w\mbox{ i.e., }C_T<\w.\end{cases}$$
This further shows that the inequality $d_{\min}^{(T)}(G)\leq d_{\min}^{(t_T)}(G')$ follows from the extended Singleton bound (see Theorem \ref{thm_extended_Singleton}). So it suffices to prove $d_{\min}^{(T)}(G)\geq d_{\min}^{(t_T)}(G').$

{\bf \textit{Case 1:}} If  $\dim(\Phi(t, G))=\Rank(F_T)=C_T<\w$, it is evident that $d_{\min}^{(T)}(G)\geq 1$.

{\bf \textit{Case 2:}} Otherwise $\dim(\Phi(t, G))=\Rank(F_T)=\w$. Notice that $\row_T(e,G')=\bzero$ for any $e\in E'\backslash E$, similarly as no path exists from $e$, $e\in E'\backslash E$, to each channel $d$, $d\in In(T)$. It follows that
\begin{align*}
d_{\min}^{(T)}(G)
=&\min\{ |\p|: \p\subseteq E \mbox{ and } \Delta(T,\p,G)\cap\Phi(T,G)\neq \{\bzero\}\}\\
=&\min\{ |\p|: \p\subseteq E \mbox{ and } \Delta(T,\p,G')\cap\Phi(T,G')\neq \{\bzero\}\}\\
=&\min\{ |\p|: \p\subseteq E' \mbox{ and } \Delta(T,\p,G')\cap\Phi(T,G')\neq \{\bzero\}\}.
\end{align*}
Define two sets of error patterns $\Pi_1$ and $\Pi_2$ as follows:
\begin{align*}
\Pi_1&=\Big\{ \p\subseteq E':\ \Delta(T,\p,G')\cap\Phi(T,G')\neq \{\bzero\}\Big\},\\
\Pi_2&=\Big\{ \p\subseteq E':\ \Delta(t_T,\p,G')\cap\Phi(t_T,G')\neq \{\bzero\}\Big\}.
\end{align*}
And for each channel $e\in E'$, let
$$\row_T(e,G')=[L_{e,d}:\ d\in In(T)].$$
For any $\p\in \Pi_1$, let $r\triangleq [r_d:\ d\in In(T)]$ be a nonzero vector in $\Delta(T,\p,G')\cap\Phi(T,G')$. Since $r\in \Delta(T,\p,G')$, there exist coefficients $a_e\in \mF$ for $e\in \p$, not all $0$, such that
\begin{align}
r&=[r_d:\ d\in In(T)]\nonumber\\
&=\sum_{e\in \p}a_e\cdot \row_T(e,G')\nonumber\\
&=\sum_{e\in \p}a_e\cdot [L_{e,d}:\ d\in In(T)]\nonumber\\
&=\begin{bmatrix}\sum_{e\in \p}a_eL_{e,d}: & d\in In(T)\end{bmatrix}.\label{equ_1}
\end{align}
On the other hand, as $r\in \Phi(T,G')$, similarly, there exist coefficients $b_e\in \mF$ for $e\in In(s)$, not all $0$, such that
\begin{align}
r&=[r_d:\ d\in In(T)]\nonumber\\
&=\sum_{e\in In(s)}b_e\cdot \row_T(e,G')\nonumber\\
&=\sum_{e\in In(s)}b_e\cdot [L_{e,d}:\ d\in In(T)]\nonumber\\
&=\begin{bmatrix}\sum_{e\in In(s)}b_eL_{e,d}:\ d\in In(T)\end{bmatrix}.\label{equ_2}
\end{align}
Combining (\ref{equ_1}) and (\ref{equ_2}), one has for all $d\in In(T)$,
$$r_d=\sum_{e\in \p}a_eL_{e,d}=\sum_{e\in In(s)}b_eL_{e,d}.$$

Further, define a row vector:
$$r'\triangleq \begin{bmatrix} \sum_{d\in In(T)}r_dk_{d,d'}:& d'\in In(t_T) \end{bmatrix},$$
where $k_{d,d'}$ is local encoding coefficient on $G'$ for the adjacent pair $(d,d')$ of channels, and $k_{d,d'}=0$ otherwise.
Then
\begin{align*}
r'&=\begin{bmatrix} \sum_{d\in In(T)}\sum_{e\in \p}a_eL_{e,d}k_{d,d'}:& d'\in In(t_T) \end{bmatrix}\\
  &=\sum_{e\in \p}a_e\begin{bmatrix} \sum_{d\in In(T)}L_{e,d}k_{d,d'}:& d'\in In(t_T) \end{bmatrix}\\
  &=\sum_{e\in \p}a_e\cdot \row_{t_T}(e,G'),
\end{align*}
and also,
\begin{align*}
r'&=\begin{bmatrix} \sum_{d\in In(T)}\sum_{e\in In(s)}b_eL_{e,d}k_{d,d'}:& d'\in In(t_T) \end{bmatrix}\\
  &=\sum_{e\in In(s)}b_e\begin{bmatrix} \sum_{d\in In(T)}L_{e,d}k_{d,d'}:& d'\in In(t_T) \end{bmatrix}\\
  &=\sum_{e\in In(s)}b_e\cdot \row_{t_T}(e,G')\neq\bzero,
\end{align*}
where the last step follows because $\row_{t_T}(e,G'), e\in In(s)$, are linearly independent, and not all of $b_e, e\in \p$, are $0$.
This means that $\Delta(t_T,\p,G')\cap \Phi(t_T,G')\neq \{\bzero\}$. Note that the result follows for any $\p\in \Pi_1$. This shows that $\p\in \Pi_2$ for any $\p\in \Pi_1$, that is, $\Pi_1\subseteq \Pi_2$.
Therefore,
$$d_{\min}^{(T)}(G)=\min_{\p\in\Pi_1}|\p|\geq \min_{\p\in\Pi_2}|\p|=d_{\min}^{(t_T)}(G')=\de_T+1.$$

Combining the two cases, we derive that the inequality (\ref{Singleton_T}) achieves with equality.
In other words, $\{ \ti{f}_e^{(In(s)\cup E)}:\ e\in E \}$ constitutes a global description of an $\w$-dimensional LNEC dispersion MDS code on $G$.

At last, the remaining problem is to determine the size of the base field. Theorem \ref{thm_e_b} shows that
$$|\mF|>\sum_{t_T\in V': C_{t_T}\geq \w}|R_{t_T}(\de_{t_T},G')|+|\{ t_T\in V': C_{t_T}<\w \}|$$
is enough. Furthermore, notice that
\begin{align*}
\sum_{t_T\in V': C_{t_T}\geq \w}|R_{t_T}(\de_{t_T},G')|=&\sum_{T\in \mT: C_T\geq \w}|R_{t_T}(\de_{T},G')|
\end{align*}
and
$$|\{ t_T\in V': C_{t_T}<\w \}|=|\{ T\in \mT: C_T<\w \}|=|V_3|.$$
Therefore, if
$$|\mF|>\sum_{T\in \mT: C_T\geq \w}|R_{t_T}(\de_{T},G')|+|V_3|,$$
there exists an $\w$-dimensional $\mF$-valued dispersion MDS code on the network $G$, which proves the theorem.
\end{IEEEproof}

Furthermore, the following corollary gives the looser lower bounds on the field size for the existence of multicast/broadcast/dispersion MDS codes.

\begin{cor}\label{cor_field_size}
Let $G=(V,E)$ be a single source acyclic network. There exists an $\w$-dimensional LNEC multicast/broadcast/dispersion MDS code on $G$, if the field sizes respectively satisfy:
\begin{align*}
|\mF|>\sum_{t\in V:\ C_t\geq \w}{|E|\choose \de_t}, \qquad |\mF|>\sum_{t\in V:\ C_t\geq \w}{|E|\choose \de_t}+|V_2|,
\end{align*}
and
\begin{align*}
|\mF|>\sum_{T\in \mT: C_T\geq \w}{|E|+\sum_{T\in \mT}\sum_{t\in T}C_t\choose \dt_T}+|V_3|,
\end{align*}
where $V_2\subseteq V$ is the set of all non-source nodes $t\in V$ with $C_t<\w$, and $V_3 \subseteq V$ is the set of all collections $T\in \mT$ with $C_T<\w$.
\end{cor}

Generally speaking, the upper bounds in Corollary \ref{cor_field_size} is larger than the corresponding one in Theorems \ref{thm_e_m} \ref{thm_e_b} and \ref{thm_e_d}, and much larger in some cases similar to the example \cite[Example 1]{Guang-MDS}.

\section{Linear Network Error Correction Generic Codes}

In this section,  we will focus on linear network error correction generic codes and the related problems.

\subsection{Notation and Definitions}

First, we require some notation and definitions similar to those ones in the last section.

\begin{defn}
In LNEC codes, let  $\xi\subseteq E$ be a nonempty collection of channels. The decoding matrix $\ti{F}_{\xi}$ at $\xi$ is defined as $\ti{F}_{\xi}=\begin{bmatrix}\ti{f}_e:& e\in \xi \end{bmatrix}$. Similarly use $\row_{\xi}(d)$ to denote the row vector of $\ti{F}_{\xi}$ indicated by the channel $d\in In(s)\cup E$. Let $F_{\xi}=\begin{bmatrix} \row_{\xi}(d_i'): & 1\leq i \leq \w \end{bmatrix}$ and
$G_{\xi}=\begin{bmatrix} \row_{\xi}(e): & e\in E\end{bmatrix}$ be two matrices of sizes $\w\times |\xi|$ and $|E| \times |\xi|$, respectively. Then
$\ti{F}_{\xi}=\begin{bmatrix} F_{\xi}\\ G_{\xi} \end{bmatrix}$.
\end{defn}

For any nonempty collection $\xi\subseteq E$ of channels, the following two vector spaces are still important.
\begin{defn}
Let $\p\subseteq E$ be an arbitrary error pattern. Define the following two vector spaces:
\begin{align*}
\Phi(\xi,G)=&\langle\{ \row_{\xi}(d_i'):\ 1\leq i\leq \w \}\rangle\\
           =&\{ (\bX \ \bzero) \cdot \ti{F}_{\xi}:\ \mbox{all $\w$-dimensional row vectors $\bX \in \mF^{\w}$} \},
\end{align*}
and
\begin{align*}
\Delta({\xi},\p,G)=&\langle\{ \row_{\xi}(e):\ e\in \p \}\rangle\\
=&\{ (\bzero \ \bZ) \cdot\ti{F}_{\xi}:\ \mbox{all $\bZ \in \mF^{|E|}$ matching error pattern $\p$} \},
\end{align*}
which is called the message space of $\xi$, and the error space of error pattern $\p$ with respect to $\xi$, respectively.
\end{defn}

As before, when there is no ambiguity, $G$ in the above notation is usually omitted and not omitted if necessary.

\begin{defn}
We say that an error pattern $\p_1$ is dominated by another error pattern $\p_2$ with respect to a nonempty collection ${\xi}\subseteq E$ of channels, if $\Delta(\xi,\p_1)\subseteq \Delta(\xi,\p_2)$ for any LNEC code. This relation is denoted by $\p_1 \prec_{\xi} \p_2$.
\end{defn}

\begin{defn}
The rank of an error pattern $\p$ with respect to a nonempty collection $\xi \subseteq E$ of channels is defined as:
$$rank_{\xi}(\p)=\min\{ |\p'|:\ \p\prec_{\xi} \p'\}.$$
\end{defn}

For the concept of rank of an error pattern, we give the following lemma which is closely similar to Lemma \ref{lem_rank}.

\begin{lemma}\label{lem_rank}
For an error pattern $\p$, introduce a source node $s_\p$. Let $\p=\{e_1,e_2,\cdots,e_l\}$, where $e_j\in In(i_j)$ for $1 \leq j \leq l$, and define $l$ new edges $e_j'=(s_\p, i_j)$. Replace each $e_j$ by $e_j'$ on the network $G$, that is, add $e_1',e_2',\cdots,e_l'$ on the network and delete $e_1,e_2,\cdots,e_l$ from the network. Then the rank of the error pattern $\p$ with respect to the collection $\xi$ in the original network is equal to the minimum cut capacity between $s_\p$ and $\xi$.
\end{lemma}

When we consider both information transmission and error correction, several definitions below are introduced.

\begin{defn}
an $\w$-dimensional LNEC code is called channel-regular, if the following holds for any nonempty collection $\xi \subseteq E$ of channels:
$$\dim(\Phi(\xi,G))=\min\{\w,C_{\xi}\}.$$
\end{defn}

If an $\w$-dimensional channel-regular LNEC code has the above property, we say that it is a LNEC generic code. If we further consider the error correction capability of a LNEC generic code, the following minimum distance are of importance.

\begin{defn}
The minimum distance of a LNEC generic code on $G$ at any nonempty collection $\xi \subseteq E$ of channels is defined as:
\begin{align*}
d_{\min}^{(\xi)}(G)&=\min\{ rank_{\xi}(\p): \Delta(\xi,\p,G)\cap\Phi(\xi,G)\neq\{\bzero\} \}\\
                   &=\min\{ |\p|: \Delta(\xi,\p,G)\cap\Phi(\xi,G)\neq\{\bzero\} \}\\
                   &=\min\{ \dim(\Delta(\xi,\p,G)): \Delta(\xi,\p,G)\cap\Phi(\xi,G)\neq\{\bzero\} \}.
\end{align*}
\end{defn}

The above minimum distance $d_{\min}^{(\xi)}(G)$ fully characterize the error-detecting and error-correcting capabilities at the nonempty collection $\xi \subseteq E$ of channels. Next, we propose the corresponding Singleton bound on $d_{\min}^{(\xi)}(G)$, called strongly extended Singleton bound.

\begin{thm}[Strongly Extended Singleton Bound]\label{thm_generic_Singleton}
For an arbitrary channel-regular linear network error correction code on an acyclic network $G=(V,E)$, let $d_{\min}^{(\xi)}(G)$ be the minimum distance at the nonempty collection $\xi\subseteq E$ of channels, and then
\begin{align}\label{Singleton_xi}
d_{\min}^{(\xi)}(G)\leq
\begin{cases}
\delta_{\xi}+1 & \mbox{ if } C_{\xi}\geq \w,\\
1              & \mbox{ if } C_{\xi} < \w.
\end{cases}
\end{align}
\end{thm}

The proof of Theorem \ref{thm_generic_Singleton} is analogous to that of Theorem \ref{thm_extended_Singleton}, so the details are omitted. And if a LNEC generic code satisfies the inequalities (\ref{Singleton_xi}) with equality, we say it a LNEC generic MDS code, or generic MDS code for short. Mathematically, for any nonempty collection $\xi\subseteq E$ of channels,
$$\dim(\Phi(\xi,G))=\Rank(F_{\xi})=\min\{\w,C_{\xi}\},$$
and
$$d_{\min}^{({\xi})}(G)=\begin{cases}\dt_{\xi}+1& \mbox{ if } C_{\xi}\geq \w,\\1& \mbox{ if } C_{\xi}< \w.\end{cases}$$

In Cai \cite{Cai}, they proposed strongly generic linear network codes and discussed several applications, particularly, including that network error correction is possible by applying strongly generic linear network codes when some conditions are satisfied. However, the analysis is briefly and incompletely. Specifically, although strongly generic linear network codes can correct network errors, the conditions of its definition is too strong for network error correction. Moreover, our refined Singleton bound (Theorem \ref{thm_generic_Singleton}) on LNEC generic codes are different from the classical case proposed by Yeung and Cai \cite{Yeung-Cai-correct-1} which is used in \cite{Cai}. In fact, a LNEC generic MDS code that satisfies the strongly extended Singleton bound with equality can correct more errors than optimal strongly generic linear network code introduced by Cai \cite[Corollary 4.5]{Cai}. Further, compared with that work, in the following, we not only give a proof of the existence of generic MDS codes but also propose an efficient algorithm for constructing it by modifying Algorithm 1 in \cite{Guang-MDS}. And an upper bound on the field size for the existence of generic MDS codes is determined.

\subsection{The Existence of LNEC Generic MDS Codes}

Kwok and Yeung \cite{Yeung-relation-dispersion-generic} indicated that, applying the existence of linear dispersion, the existence of generic codes can be obtained when the size of the base field $\mF$ is sufficiently large. In this subsection, we also give a positive answer to the existence question of generic MDS codes, which is proved by applying the existence of dispersion MDS codes. But in this case, the proof is more complicated but more interesting, and involves more technical arguments.

\begin{thm}\label{thm_e_g}
Let $G=(V,E)$ be an acyclic network. There exists an $\w$-dimensional $\mF$-valued LNEC generic MDS code on $G$, if the size of the base field satisfies:
\begin{align*}
|\mF|>\sum_{T\in \mT: C_T\geq \w \mbox{ \footnotesize in }G'}{2|E|+\sum_{T\in \mT}\sum_{t\in T}C_t \choose \de_T}+|\{T\in \mT: C_T<\w \mbox{ in }G'\}|,
\end{align*}
where $\mT$ is the set of all collections of non-source nodes in the new network $G'$ corresponding to $G$, as described at the beginning of the proof.
\end{thm}

\begin{IEEEproof}
From the acyclic network $G$, we need to construct a new network $G'=(V',E')$ as follows: for each channel $e=(i,j)\in E$, install a new node $n_e$ between $i$ and $j$, and $n_e$ splits the channel $e$ into two channels $e_1=(i,n_e),\ e_2=(n_e,j)$\footnote{This method is mentioned in Section \ref{Sec_preliminaries}, in order to determine the minimum cut capacity between the source node $s$ and a nonempty collection $\xi$ of channels.}.

Thus, it is not difficult to see that $V'=V \cup \{n_e: e\in E\}$ and $E'=E_1\cup E_2$, where $E_i=\{e_i: e\in E\}$, $i=1,2$, which easily shows that
$$|V'|=|V \cup \{n_e: e\in E\}|=|V|+|E|$$
and
$$|E'|=|E_1\cup E_2|=2|E|.$$

Next, in the network $G'$, we always use $e_1$ and $e_2$ to denote the channels $(tail(e),n_e)$ and $(n_e,head(e))$ in $G$, respectively. Recall that for the network $G$, there is an upstream-to-downstream order among all channels in $E$. Assume that $E=\{e_k: 1\leq k\leq |E|\}$ and without loss of generality, assume that the order is
$$e_1 \prec e_2\prec \cdots \prec e_{|E|}.$$
We further extend this order to the new network $G'$:
$$e_{1,1} \prec e_{1,2} \prec e_{2,1} \prec e_{2,2} \prec \cdots \prec e_{|E|,1} \prec e_{|E|,2}$$
and it is easy to check that this order is also upstream-to-downstream in $E'$.

For this new network $G'$, by Theorem \ref{thm_e_d}, there exists a LNEC dispersion MDS code for some finite field $\mF$. Let $\{ k_{d,e}(G'):d,e\in In(s)\cup E' \}$ and $\{ \ti{f}_e(G'): e\in E' \}$ be the local description and the global description of this LNEC code, respectively, where note that $\ti{f}_e(G')$ is an $(\w+2|E|)$-dimensional column vector for any channel $e\in E'$. For any channel $e\in E$ in $G$, define an $(\w+|E|)$-dimensional column vector
\begin{align}\label{f_e_generic}
\ti{f}_e(G)\triangleq \ti{f}_{e_1}^{(In(s)\cup E_1)}(G'),
\end{align}
which is formed by those entries indexed by all channels in $In(s)\cup E_1$. Next, we will show that $\{ \ti{f}_e(G): e\in E \}$ constitutes a global description of a LNEC generic MDS code on $G$.

First, we verify that $\{ \ti{f}_e(G): e\in E \}$ constitutes a global description of a LNEC code on $G$. In other words, we need to indicate that for all $e\in E$, the following recursive formulae hold:
$$\ti{f}_e(G)=\sum_{d\in In(tail(e)) \text{ in } G}k_{d,e}(G)\ti{f}_d(G)+1_e(G)$$
with the boundary conditions
$\ti{f}_{d_i'}(G)=1_{d_i'}(G)$, $1\leq i \leq \w$, where $k_{d,e}(G)\in \mF$ is the local encoding coefficient on $G$ and $1_e(G)$ is an $(\w+|E|)$-dimensional column vector which is the indicator function of $e\in In(s)\cup E$. For the given LNEC code on $G'$, we have for any $e_1\in E_1$,
\begin{align*}
\ti{f}_{e_1}(G')=&\sum_{d\in In(tail(e_1)) \text{ in } G'}k_{d,e_1}(G')\ti{f}_d(G')+1_{e_1}(G')\\
=&\sum_{d\in In(tail(e)) \text{ in } G}k_{d_2,e_1}(G')\ti{f}_{d_2}(G')+1_{e_1}(G')\\
=&\sum_{d\in In(tail(e)) \text{ in } G}k_{d_2,e_1}(G')[k_{d_1,d_2}(G')\ti{f}_{d_1}(G')+1_{d_2}(G')]+1_{e_1}(G').
\end{align*}
Consequently,
\begin{align*}
\ti{f}_{e_1}^{(In(s)\cup E_1)}(G')=\sum_{d\in In(tail(e)) \text{ in } G}k_{d_2,e_1}(G')k_{d_1,d_2}(G')\ti{f}_{d_1}^{(In(s)\cup E_1)}(G')+1_{e_1}^{(In(s)\cup E_1)}(G').
\end{align*}
Let $k_{d,e}(G)=k_{d_2,e_1}(G') k_{d_1,d_2}(G')$, and note that
$$1_{e_1}^{(In(s)\cup E_1)}(G')=1_{e}(G),$$
and $1_{d_2}^{(In(s)\cup E_1)}(G')$ is an $(\w+|E|)$-dimensional all-zero column vector. Thus, it follows that
$$\ti{f}_{e}(G)=\sum_{d\in In(tail(e)) \text{ in } G}k_{d,e}(G)\ti{f}_{d}(G)+1_{e}(G),$$
which means that $\{ \ti{f}_e(G): e\in E \}$ constitutes a global description of a LNEC code on $G$.

Next, we further prove that $\{ \ti{f}_e(G): e\in E \}$ is a generic MDS code on $G$. For any nonempty collection $\xi \subseteq E$ of channels, recall $\Phi(\xi,G)=\langle\{ \row_{\xi}(d_i',G): 1\leq i \leq \w \}\rangle$. In $G'$, define a collection $T_\xi$ of non-source nodes $T_{\xi}=\{ n_e: e\in \xi \}\subseteq V'$, and collections $\xi_i$ of channels $\xi_i=\{ e_i: e\in \xi \}\subseteq E_i \subseteq E'$, $i=1,2$. Thus, by the definition (\ref{f_e_generic}) of $\ti{f}_e(G)$, one has
$$\row_{\xi}(d_j',G)=\row_{T_\xi}(d_j',G')$$
for all imaginary source channels $d_j'$, $1\leq j \leq \w$, which implies that
\begin{align}\label{eq_3}
\Phi(\xi,G)=\Phi(T_\xi,G').
\end{align}
Furthermore, since the minimum cut capacity between $s$ and $\xi$ in $G$ is equal to that between $s$ and $T_{\xi}$ in $G'$, i.e., $C_{\xi}=C_{T_{\xi}}$, and the given LNEC code on $G'$ is a dispersion MDS code, it follows that
\begin{align*}
\dim(\Phi(\xi,G))=\dim(\Phi(T_\xi,G'))=\min\{ \w,C_{T_\xi} \}=\min\{ \w,C_{\xi} \}.
\end{align*}

At last, the remaining problem is the determination of the minimum distance $d_{\min}^{(\xi)}(G)$.

{\bf \textit{Case 1:}} Taking into account all nonempty collections $\xi \subseteq E$ of channels with $C_{\xi}<\w$, by strongly extended Singleton bound (see Corollary \ref{thm_generic_Singleton}), we know $d_{\min}^{(\xi)}(G)\leq 1$ for any $\xi\subseteq E$ with $C_{\xi}<\w$.
on the other hand, obviously,
$$d_{\min}^{(\xi)}(G)=\min\{|\p|: \Delta(\xi,\p,G)\cap \Phi(\xi,G)\neq \{\bzero\} \}\geq 1.$$
Combining the above, one has $d_{\min}^{(\xi)}(G)=1$ for any $\xi\subseteq E$ with $C_{\xi}<\w$.

{\bf \textit{Case 2:}} Otherwise, taking into account all nonempty collections $\xi\subseteq E$ with $C_{\xi}\geq \w$,
the definition (\ref{f_e_generic}) of global encoding kernel $\ti{f}_e(G)$ on $G$ implies, for any $e\in E$,
$$\row_{\xi}(e,G)=\row_{T_\xi}(e_1,G').$$
And thus, for any error pattern $\p\subseteq E$,
\begin{align}\label{eq_4}
\Delta(\xi,\p,G)&=\langle\{ \row_{\xi}(e,G): e\in \p \}\rangle\nonumber\\
                &=\langle\{ \row_{T_\xi}(e_1,G'): e_1\in \p_1 \}\rangle\nonumber\\
                &=\Delta(T_\xi,\p_1,G'),
\end{align}
where $\p_1\triangleq\{e_1: e\in \p\}\subseteq E_1 \subseteq E'$ is an error pattern in $G'$ corresponding to $\p$.

Define two sets of error patterns:
\begin{align*}
\Pi(\xi,G)&=\{\p\subseteq E: \Delta(\xi,\p,G)\cap \Phi(\xi,G)\neq \{\bzero\}\},\\
\Pi(T_\xi,G')&=\{\p'\subseteq E': \Delta(T_{\xi},\p',G')\cap \Phi(T_\xi,G')\neq \{\bzero\}\}.
\end{align*}
Thus one obtains
$$d_{\min}^{(\xi)}(G)=\min_{\p\in \Pi(\xi,G)}|\p|,\quad \mbox{and}\quad d_{\min}^{(T_\xi)}(G')=\min_{\p\in \Pi(T_\xi,G')}|\p|.$$
For any error pattern $\p\in \Pi(\xi,G)$, let $\p_1=\{e_1: e\in \p\}\subseteq E_1\subseteq E'$.
It follows from (\ref{eq_3}) and (\ref{eq_4}) that
$$\Delta(T_\xi,\p_1,G')\cap \Phi(T_\xi,G')
=\Delta(\xi,\p,G)\cap \Phi(\xi,G)\neq \{\bzero\}.$$
This leads to $\p_1\in \Pi(T_\xi,G')$, which, together with the fact $|\p_1|=|\p|$, gives that
\begin{align*}
d_{\min}^{(\xi)}(G)=\min_{\p\in \Pi(\xi,G)}|\p|\geq\min_{\p'\in \Pi(T_\xi,G')}|\p'|=d_{\min}^{(T_\xi)}(G').
\end{align*}
Further, we know
$$d_{\min}^{(T_\xi)}(G')=C_{T_\xi}-\w+1=C_{\xi}-\w+1=\de_{\xi}+1,$$
which shows $d_{\min}^{(\xi)}(G)\geq \de_{\xi}+1$.

On the other hand, by strongly extended Singleton bound, one obtains $d_{\min}^{(\xi)}(G)\leq \de_\xi+1$. Therefore, we deduce $d_{\min}^{(\xi)}(G)=\de_\xi+1$. At last, the field size for the existence of generic MDS codes is taken into account. It is not difficult to see that Corollary \ref{cor_field_size} provides a bound on the field size, that is,
\begin{align*}
|\mF|>\sum_{T\in \mT: C_T\geq \w \mbox{ \footnotesize in }G'}{2|E|+\sum_{T\in \mT}\sum_{t\in T}C_t \choose \de_T}+|\{T\in \mT: C_T<\w \mbox{ in }G'\}|.
\end{align*}
This completes the proof.
\end{IEEEproof}

The following theorem indicates the relation between generic MDS codes and dispersion MDS codes.

\begin{thm}
Every LNEC generic MDS code on $G$ is a LNEC dispersion MDS code on $G$.
\end{thm}
\begin{IEEEproof}
Let $T$ be an arbitrary collection of non-source nodes, and let $\xi=In(T)=\cup_{t\in T}In(t)$. It is readily seen that $C_{\xi}=C_T$. Further, since this LNEC generic MDS code on $G$ is channel-regular, it follows that
$$\dim(\Phi(\xi))=\min\{\w,C_{\xi}\}=\min\{\w,C_T\}.$$
Together with
\begin{align*}
\Phi(\xi)=\langle\{ \row_{\xi}(d_i'): 1\leq i \leq \w \}\rangle=\langle\{ \row_T(d_i'): 1\leq i \leq \w \}\rangle=\Phi(T),
\end{align*}
this means that $\dim(\Phi(T))=\min\{\w,C_T\}$, that is, this LNEC code is strongly sup-regular.
Subsequently,
\begin{align*}
\Delta(\xi,\p)=\langle\{ \row_{\xi}(e): e\in\p \}\rangle=\langle\{ \row_T(e): e\in\p \}\rangle=\Delta(T,\p).
\end{align*}
And thus
\begin{align*}
d_{\min}^{(\xi)}&=\min\{ |\p|:\Delta(\xi,\p)\cap \Phi(\xi)\neq \{\bzero\} \}=\min\{ |\p|: \Delta(T,\p)\cap \Phi(T)\neq \{\bzero\} \}=d_{\min}^{(T)}.
\end{align*}
Together with the definition of generic MDS codes, one also derives:
\begin{align*}
d_{\min}^{(T)}(G)=d_{\min}^{({\xi})}(G)&=
\begin{cases}C_{\xi}-\w+1& \mbox{ if } C_{\xi}\geq \w\\1& \mbox{ if } C_{\xi}< \w\end{cases}\\
&=\begin{cases}C_{T}-\w+1& \mbox{ if } C_T\geq \w\\1& \mbox{ if } C_T< \w . \end{cases}
\end{align*}
This completes the proof.
\end{IEEEproof}

This theorem and Remark \ref{rem_relation-m-b-d} show that every LNEC generic MDS code is also a LNEC multicast/broadcast MDS code. But a multicast MDS code is not necessarily a broadcast MDS code, a broadcast MDS code is not necessarily a dispersion MDS code, and a dispersion MDS code is not necessarily a generic MDS code, because, at least, a linear multicast is not necessarily a linear broadcast, a linear broadcast is not necessarily a linear dispersion, and a linear dispersion is not necessarily a network generic code.


\section{Construction of LNEC Multicast/Broadcast/Dispersion/Generic MDS Codes}

By the proofs of Theorems \ref{thm_e_b}, \ref{thm_e_d} and \ref{thm_e_g}, if we can design an algorithm for constructing multicast MDS codes, then it is not difficult to obtain algorithms for constructing broadcast MDS codes, dispersion MDS codes and generic MDS codes. In this section, by modifying the algorithm for constructing ordinary MDS codes, i.e., \cite[Algorithm 1]{Guang-MDS}, we give the following Algorithm \ref{algo_m} for constructing multicast MDS codes. First, we introduce some notation.
For arbitrary subset $B\subseteq In(s)\cup E\cup E'$, where $E'$ is the set of all imaginary error channels on $G$, define
\begin{align*}
\tilde{\mL}(B)=&\langle \{ \f_e: e\in B \} \rangle,\qquad \tilde{\mL}^{\p}(B)=\langle \{ \f^{\p}_e: e\in B \} \rangle,\\
\mL^{\p}(B)=&\langle \{ f^{\p}_e: e\in B \} \rangle,\qquad  \mL^{\p^c}(B)=\langle \{ f^{\p^c}_e: e\in B \} \rangle.
\end{align*}

\begin{algorithm}[!htp]\label{algo_m}
\SetAlgoLined
\KwIn{The single source acyclic network $G=(V,E)$, and the information rate $\w$.}
\KwOut{Extended global encoding kernels (forming a global description of a multicast MDS code).}
\KwInitialization{
\begin{enumerate}\rm
  \item For each non-source node $t\in V$ with $C_t\geq \w$ and each $\p\in R_t(\dt_t)$, find $(\w+\dt_t)$ channel-disjoint paths from $In(s)$ or $\p'$ to $t$ satisfying Lemma {\ref{lem_path}}, where $\p'$ is the set of imaginary error channels $e'$ corresponding to $e\in \p$, i.e., $\p'=\{e': e\in \p \}$.\\
        Denote by $\mP_{t,\p}$ the set of the chosen $(\w+\dt_t)$ channel-disjoint paths, and $E_{t,\p}$ denotes the set of all channels on paths in $\mathcal{P}_{t,\p}$;
  \item For each non-source node $t\in V$ with $C_t\geq \w$ and each $\p\in R_t(\dt_t)$, initialize dynamic channel set $CUT_{t,\p}=In(s)\cup\p'=\{d_1',\cdots,d_\w'\}\cup\{e': e\in\p\},$
      as well as the extended global encoding kernels $\f_e=1_e$ for all imaginary channels $e\in In(s)\cup E'$.
\end{enumerate}}
\Begin{
\ForEach{ {\rm node $i\in V$ (according to the upstream-to-downstream order of nodes)}}{
  \ForEach{ {\rm channel $e\in Out(i)$ (according to an arbitrary order)}}{
    \eIf{$e\notin \cup_{t\in V:\ C_t\geq \w}\cup_{\p\in R_t(\dt_t)}E_{t,\p}$,}{
       $ \tilde{f}_e=1_e$\;
       all $CUT_{t,\p}$ remain unchanged.}{
    choose $\g_e\in\tilde{\mL}(In(i)\cup\{e'\})\backslash\cup_{t\in V:\atop C_t\geq \w}\cup_{\p\in R_t(\dt_t):\atop e\in E_{t,\p}}[\mL^{\p}(CUT_{t,\p}\backslash \{e(t,\p)\})+\mL^{\p^c}(In(i)\cup \{e'\})]$, where $e(t,\p)$ represents the previous channel of $e$ on the path which $e$ locates on.\\
       \eIf{$\g_e(e)=0$}{
          $\f_e=\g_e+1_e$\;}{
       $\f_e=\g_e(e)^{-1}\cdot\g_e$.}
       For those $CUT_{t,\p}$ satisfying $e\in E_{t,\p}$, update $CUT_{t,\p}=\{CUT_{t,\p}\backslash\{e(t,\p)\}\}\cup\{e\}$\; for others, $CUT_{t,\p}$ remain unchanged.
       }
}
}
}
\caption{The algorithm for constructing multicast MDS codes.}
\end{algorithm}

\begin{rem}\
\begin{itemize}
  \item The algorithm verification is similar to that of {\rm \cite[Algorithm 1]{Guang-MDS}}, and in {\rm \cite{Guang-MDS}} the authors gave the detailed verification of the algorithm, so it is omitted. Actually, the proposed algorithm can also construct a general LNEC multicast (resp. broadcast, dispersion, and generic) code just by replacing the redundancy $\de_t$ by any positive $\bt$ with $\bt\leq \de_t$ for each $t\in V$, and so the constructed LNEC multicast code has the property $d_{\min}^{(t)}\geq \bt$ for the node $t$.
  \item Next, from \cite{Guang-MDS} it is not difficult to analyze the time complexity of the algorithm. If we use the method of Testing Linear Independent Quickly \cite[\Rmnum{3},A]{co-construction} (briefly speaking, choose a vector randomly, and then test its linear independence on other vectors), the expected time complexity of the algorithm is at most:
\begin{align*}
\mO\Big(|E|\sum_{t\in V:\atop C_t\geq \w}C_t\cdot\left[{ |E|\choose \de_t}+|R_t(\de_t)|(\w+\frac{|E|+1}{2})\right] \Big).
\end{align*}
If we use the method of Deterministic Implementation \cite[\Rmnum{3},B]{co-construction} (briefly speaking, use a deterministic method for choosing a vector which is linear independence on other vectors), the total time complexity of the algorithm is at most:
\begin{align*}
\mO\Big( |E|\sum_{t\in V:\atop C_t\geq \w}C_t\cdot\Big[{|E|\choose \de_t}+|R_t(\de_t)|\bigg(\sum_{t\in V:\atop C_t\geq \w} |R_t(\de_t)|+C_t\bigg)\Big]\Big).
\end{align*}
  \item Further, Algorithm \ref{algo_m} can also imply the fact below: when we consider constructing these types of MDS codes by random method, i.e., choosing the local coefficients $k_{d,e}$ for all adjacent pairs  $(d,e)$ of channels independently according to the uniform distribution on the base field $\mF$, these four classes of MDS codes can be constructed with probability tending to $1$ as $|\mF|\rightarrow \infty$.
\end{itemize}
\end{rem}


\section{Conclusion}
In linear network coding, Yeung {\it et al.} \cite{Zhang-book} \cite{Yeung-book} define four important classes of linear network codes: linear multicast, linear broadcast, linear dispersion, and generic network codes. More generally, when channels of networks are noisy, information transmission and error correction have to be under consideration simultaneously, and thus we further proposed four classes of LNEC codes regarded as the generalizations of the original four classes of linear network codes, that is, LNEC multicast/broadcast/dispersion/generic codes. Furthermore, we present the extended Singleton bounds for them. Obviously, it is expected to apply those LNEC codes achieving the extended Singleton bounds with equality. So, we further define the corresponding optimal codes, that is, LNEC multicast/broadcast/dispersion/generic MDS codes, and show their existence. Finally, explicit constructions of such codes are proposed.

Some interesting problems in this direction remain open. For instance, similar to the variable-rate LNEC MDS codes studied in \cite{Guang-uni-MDS},
we also want to design the corresponding variable-rate MDS codes for these MDS codes for practical applications. For non-coherent networks, we have to consider constructing these new types of LNEC codes by random method. And thus we care about its performance. In other words, the same problems in random linear network coding would be discussed.

\section*{Acknowledgment}
The authors would like to thank Prof. Zhen Zhang for his insightful comments and helpful discussions on the topic of this paper.


\end{document}